\documentclass[pra,twocolumn,superscriptaddress,floatfix]{revtex4-1}

\usepackage{amsmath,amssymb} 
\usepackage{bm} 
\usepackage{graphicx} 
\usepackage{comment} 
\usepackage{textcomp} 
\usepackage{braket}

\usepackage{enumitem}
\setlist{noitemsep,leftmargin=*,topsep=0pt,parsep=0pt}

\usepackage{xcolor} 
\definecolor{lightgray}{gray}{0.6}
\definecolor{medgray}{gray}{0.4}

\newcommand{\be}{\begin{equation}}
\newcommand{\ee}{\end{equation}}
\newcommand{\bea}{\begin{eqnarray}}
\newcommand{\eea}{\end{eqnarray}}

\usepackage{hyperref}
\hypersetup{
colorlinks=true,
urlcolor= blue,
linkcolor= blue,
}
\usepackage[maxfloats=256]{morefloats}
\newif\ifptitle
\newif\ifpnumber
\newcounter{para}

\ptitletrue  
\pnumbertrue  

\begin{document}

\title {Topology and Polarization of Optical Vortex Fields from Atomic Phased Arrays
}

\author{Hao Wang$^1$ and Andrei Afanasev}

\affiliation{Department of Physics,
The George Washington University, Washington, DC 20052, USA}

\begin{abstract}
We developed theoretical formalism for generation of optical vortices by phased arrays of atoms. Using Jacobi-Anger expansion, we demonstrate the resulting field topology and determine the least number of array elements necessary for generation of vortices with a given topological charge. Vector vortices were considered, taking into account both spin and orbital angular momenta of electromagnetic fields. It was found that for the vortex field near the phase singularity, the transverse-position dependence of 3D polarization parameters is independent of the distance to the radiation source. 
\end{abstract}
\date{\today
}
\maketitle


\section{Introduction}

Studies and applications of the optical angular momentum (OAM) beams are widespread, and the methods for generation of optical vortices are well developed \cite{franke2017light}. Of particular interest are enhanced capabilities of OAM beams to transmit information encoded in the orthogonal angular-momentum eigenstates \cite{Willner2021oam}. OAM radio have a potential to improve the bandwidth of radio communications \cite{thide2021oam}.

Optical vortices can be formed using coherent laser beam arrays \cite{WANG20091088} or from an array of small apertures with appropriate phases Ref.\cite{Liu_2015}. Our work was motivated by the studies of topological states of two-dimensional atomic phased arrays \cite{perczel2017topological} in the context of quantum computing and transfer of quantum information.  

Given a limited number of atoms in an array, a question arises of what number of atoms is required to generate radiation - or a vortex - with a certain topological charge. In the previous study \cite{Wang_2022}, our consideration was limited to scalar fields and 2D-arrays of point-like emitters. In the present paper, we extend our formalism to 3D vector fields of phased atomic antenna array and analyze topological and polarization properties of the resulting radiation.

The paper is organized as follows. In Sec.IV, we analyze a radiation from a circular array of polarized atoms and obtain an analytic formula for the case of large number of atoms. In Sec.III, the case of a finite number of atoms is treated via Jacobi-Anger expansion that enables us to perform analytic summation over all individual emitters. Sec.IV provides a review of the formalism of 3D polarization of optical fields and gives analytic results of polarization parameters in limiting cases. Sect.V gives a detailed numerical analysis of atomic-array radiation intensity and polarization as a function of the propagation distance and a number of radiating atoms. The results are summarized in Sec.VI. 

\section{\label{sec:Start}Radiation of an atomic phased array with many elements}

\begin{figure}[h]
\centering
\includegraphics[width=0.35\textwidth]{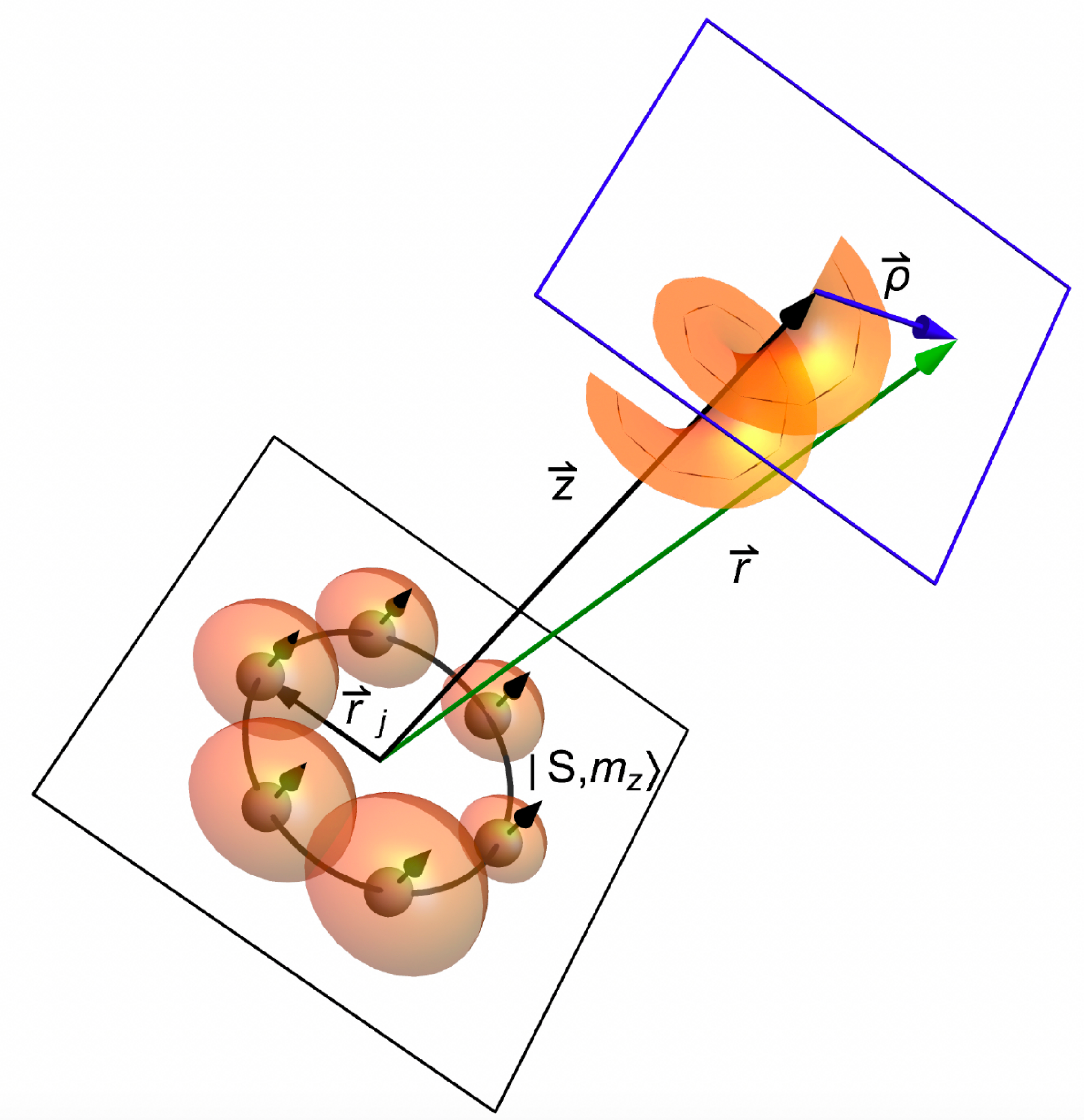}
\caption{Phased atomic ring array with atoms polarized along $z$-axis in spin states $\ket{S,m_z}$. Location in the parallel observation plane at a distance $z$ is defined by the vector $\vec \rho$. A vector $\vec r_j$ defines position of the $j$-th atom, $|\vec r_j|=R$. Shown schematically are spherical wave fronts of radiation from individual atoms and a helical wave front of the the field resulting from their interference.
}
\label{fig:geometry}
\end{figure}

Consider a circular array of $N$ excited atoms, each carrying an orbital angular momentum $S$ and its projection $m_z$. The array geometry is shown in Fig.\ref{fig:geometry}, along with the observation plane located at a distance $z$ parallel to the atomic array. Let us denote a state vector of light emitted by the individual $j$-th atom located at a polar angle $\phi_j$ on the ring as $\ket{\phi_j,S,m_z}$. Each adjacent atom radiates with a phase difference $\Delta\phi=\frac{2\pi l}{N}$, where $l$ is an externally controlled phase parameter. The state vector of a photon emitted by the $j$-th  atom is
\begin{equation}
\label{eq:ad7a}
    \begin{split}
        \ket{\phi_j,S,m_z}\to e^{il\phi_j}\ket{m_z}_j,
    \end{split}
\end{equation}
where position dependence for the individual source is indicated by the source number $j$.

Using the photon field operator $\hat{A}^{\mu}$, the total field of the emitted light of an atomic phased ring array is
\begin{equation}
\label{eq:ad1}
    \begin{split}
        A^{\mu}_{m_z}(\vec r)=\sum_{j}e^{il\phi_j}\bra{0}\hat{A}^{\mu}\ket{m_z}_j.
    \end{split}
\end{equation}

For each atom, the polarization state of the emitted light can be expanded in the basis formed by left- and right-circularly polarized light in its propagation direction along $\vec r - \vec r_j$,

\begin{equation}
\label{eq:ad2}
    \begin{split}
        \ket{m_z}_j=\alpha^{(j)}_{+}\ket{\Lambda=+1}+\alpha^{(j)}_{-}\ket{\Lambda=-1},
    \end{split}
\end{equation}
where $\Lambda=\pm 1$ defined photon's helicity and the expansion coefficients are
\begin{equation}
\label{eq:d2}
\alpha_{\pm}^{(j)}=\  \braket{m_z|_j\Lambda=\pm1}.
\end{equation}

It should be noted that observation of the light emitted by the atom is near a point on  $z$-axis aligned with the center of the ring array. The light propagating in $\ket{\Lambda=\pm 1}$-state in the above expansion has an angle $\theta_j$ with respect to the z-axis. The angle is the same for all the atoms $\theta_j\equiv\theta_k$. In order to choose quantization along z-axis, we have to rotate the coordinate system attached to the atom by a rotation operator $R(-\phi_j,\theta_k,\phi_j)$. Therefore, the coefficients become
\begin{equation}
\label{eq:d3}
    \begin{split}
        \alpha_{+}^{(j)}= \bra{m_z}R(-\phi_j,\theta_k,\phi_j)\ket{\Lambda=+1}\\
        \alpha_{-}^{(j)}= \bra{m_z}R(-\phi_j,\theta_k,\phi_j)\ket{{\Lambda=-1}}.
    \end{split}
\end{equation}

The above expressions give Wigner $d$-matrices for the considered case of  electric dipole emission (while generalization to higher multipoles is straightforward):
\begin{equation}
\label{eq:d4}
    \begin{split}
        \alpha_{+}^{(j)}=d^1_{1 m_z}(\theta_k)e^{i(m_z-1)\phi_j}\\
        \alpha_{-}^{(j)}=d^1_{-1 m_z}(\theta_k)e^{i(m_z+1)\phi_j},
    \end{split}
\end{equation}
where $d^1_{\pm m_z}(\theta_k)=(1\pm m_z\cos\theta_k)/2$.

The total field of the emitted light of an atomic phased ring array can then be expressed in the basis formed by left- and right- circularly polarized light by the above coefficients. The  angles $\phi_j$ and $\theta_k$ define orientation of the wave vector $\vec{k}_j$ of light emitted by the j-th atom in spherical coordinate system. 

\begin{equation}
\label{eq:ad3}
    \begin{split}
        &A^{\mu} _{m_z}=\sum_{j}e^{il\phi_{j}}\\
        &(\alpha_{+}^{(j)}\bra{0}\hat{A}^{\mu}\ket{\vec{k}_j,\Lambda=+1}+\alpha_{-}^{(j)}\bra{0}\hat{A}^{\mu}\ket{\vec{k}_j,\Lambda=-1}).
    \end{split}
\end{equation}

If the number of atoms in the ring goes to infinity, it gives us the continuous case. We notice that in momentum space infinite number of states of light $\ket{\Lambda}_j$ lying on a cone with an opening angle $2\theta_k$ form a vortex state of light. Hence with infinitely many atoms,
$$
\sum_{j}\to\int\limits\frac{d\phi_j}{2\pi},
$$
the total field generated by the atomic phased ring array can be expressed by two vector vortices with a potential $A^{\mu}_{\Lambda=\pm1}$ with opposite helicities,  

\begin{equation}
\label{eq:d1}
    \begin{split}
        A^{\mu} _{m_z}=\alpha_{+}A^{\mu}_{\Lambda=+1}+\alpha_{-}A^{\mu}_{\Lambda=-1},
    \end{split}
\end{equation}
where $\alpha_{\pm}=\frac{1}{2}(1\pm m_z\cos\theta_k)$. Note that $(\alpha_-^2+\alpha_+^2)^{-1/2}=[(1+\cos(\theta_k)^2)/2]^{-1/2}$ has to be factored in for proper normalization of the field amplitude.

A vector vortex field formed by a cone of light rays with helicity $\Lambda$ can be expressed as \cite{Jentschura11,PhysRevA.88.033841}
\begin{equation}
\label{eq:d6}
    \begin{split}
        A^{\mu}_{\Lambda}=e^{-i(\omega t-k_zz)}&\sqrt{\frac{\kappa}{2\pi}}A\\
        &\int\limits\frac{d\phi_j}{2\pi}(-i)^{m_\gamma}e^{im_{\gamma}\phi_j}e^{i\vec{\kappa_j}\cdot\vec{\rho}}\vec{\epsilon}_{\Lambda},
    \end{split}
\end{equation}
$m_{\gamma}=l+\Lambda$, where $m_{\gamma}$ is a total angular momentum projection of the vortex photon, while $l$ can be interpreted as an orbital angular momentum projection of the photon in a paraxial limit $\theta_k\to 0$. The quantities  $\kappa$ and $\rho$ are the transverse components of wave vector and a position vector in an observation plane, respectively; $\vec \kappa_j\cdot \vec \rho=\vec k_j\cdot \vec\rho=\kappa \rho \cos(\phi_j)$ and $A$ is an overall normalization factor. Consider the term with $\Lambda=+1$. The unit vectors are
\begin{equation}
\label{eq:d7a}
    \begin{split}
    \epsilon^{\mu}_{\Lambda=+1}=e^{-i\phi_j}\cos^2\frac{\theta_k}{2}\eta_{+1}^{\mu}+e^{i\phi_j}\sin^2\frac{\theta_k}{2}\eta_{-1}^{\mu}+\frac{1}{\sqrt{2}}\sin\theta_k\eta_{0}^{\mu},
    \end{split}
\end{equation}
where 
\begin{equation}
\label{eq:d7}
    \begin{split}
    \eta_{\pm1}^{\mu}=\frac{1}{\sqrt{2}}\begin{pmatrix}
    0;&\mp1,&-i,&0
    \end{pmatrix}\\
    \eta_{0}^{\mu}=\frac{1}{\sqrt{2}}\begin{pmatrix}
    0;&0,&0,&1
    \end{pmatrix}
    \end{split}
\end{equation}
are the spherical-basis vectors with a $z$-axis orthogonal to the array, as shown in Fig.1.
Hence we have 
\begin{equation}
\label{eq:d8}
    \begin{split}
        A^{\mu} _{m_z}=e^{-i(\omega t-k_zz)}\sqrt{\frac{\kappa}{2\pi}}A\int\limits\frac{d\phi_j}{2\pi}e^{i\vec{\kappa_j}\cdot\vec{\rho}}\\ [\frac{1}{2}(1+m_z\cos\theta_k)e^{i(m_z-1)\phi_j}(-i)^{l+1}e^{i(l+1)\phi_j}\epsilon^{\mu}_{+1}\\+\frac{1}{2}(1-m_z\cos\theta_k)e^{i(m_z+1)\phi_j}(-i)^{l-1}e^{i(l-1)\phi_j}\epsilon^{\mu}_{-1}].
    \end{split}
\end{equation}
Substituting expressions (\ref{eq:d7}) for basis vectors,  we obtain
\begin{equation}
\label{eq:d9}
    \begin{split}
        A^{\mu} _{m_z}=e^{-i(\omega t-k_zz)}\sqrt{\frac{\kappa}{2\pi}}A\int\limits\frac{d\phi_j}{2\pi}e^{i\vec{\kappa_j}\cdot\vec{\rho}}e^{i(l+m_z)\phi_j}\\ [\frac{1}{2}(1+m_z\cos\theta_k)(-i)^{l+1}\\
        (e^{-i\phi_j}\cos^2\frac{\theta_k}{2}\eta_{1}^{\mu}+e^{i\phi_j}\sin^2\frac{\theta_k}{2}\eta_{-1}^{\mu}+\frac{1}{\sqrt{2}}\sin\theta_k\eta_{0}^{\mu})\\+\frac{1}{2}(1-m_z\cos\theta_k)(-i)^{l-1}\\(e^{i\phi_j}\cos^2\frac{\theta_k}{2}\eta_{-1}^{\mu}+e^{-i\phi_j}\sin^2\frac{\theta_k}{2}\eta_{1}^{\mu}-\frac{1}{\sqrt{2}}\sin\theta_k\eta_{0}^{\mu})].
    \end{split}
\end{equation}

The integral over $\phi_j$ can be expressed in terms of Bessel functions,
\begin{equation}
\label{eq:d10}
    \begin{split}
       \int\limits\frac{d\phi_j}{2\pi}e^{i\vec{\kappa_j}\cdot\vec{\rho}}e^{i(l+m_z-1)\phi_j}=- iJ_{l+m_z-1}(\kappa\rho)e^{i(l+m_z-1)\phi_{\rho}}.
    \end{split}
\end{equation}
As a result, a vector potential for radiation from continuously distributed atoms ($N\to \infty$) around the ring array is
\begin{eqnarray}
\label{eq:d12}
        &&A^{\mu} _{m_z}=e^{-i(\omega t-k_zz)}\sqrt{\frac{\kappa}{2\pi}}A\nonumber \\ &&(-\frac{ i}{2}e^{i(l+m_z-1)\phi_{\rho}}J_{l+m_z-1}(\kappa\rho)(1+m_z\cos^2\theta_k)\eta_{1}^{\mu}\\&&+\frac{ i}{2}e^{i(l+m_z+1)\phi_{\rho}}J_{l+m_z+1}(\kappa\rho)(1-m_z\cos^2\theta_k)\eta_{-1}^{\mu}\nonumber\\&&+\frac{m_z }{2\sqrt{2}}e^{i(l+m_z)\phi_{\rho}}J_{l+m_z}(\kappa\rho)\sin2\theta_k\eta_{0}^{\mu}).\nonumber
\end{eqnarray}

\section{Radiation field from a finite number of atomic sources}

In the discrete case, recall the expression Eq.(\ref{eq:ad1}). The light emitted by the $j$-th atom has the same angular distribution as the light emitted by an oscillating dipole \cite{sobelman2012atomic}.

\begin{equation}
\label{eq:ad4}
\begin{split}
\bra{0}\hat{A}^{\mu}\ket{m_z}_j=\frac{A_N}{16\pi}\frac{\omega_0^2e^2}{\epsilon_0c^2}\frac{1}{|\vec{r}-\vec {r}_j|}e^{ik|\vec{r}-\vec{r}_j|}e^{-i\omega t}\\
[(1+m_z\cos^2\theta_j)e^{i(m_z-1)\phi_j}\eta^{\mu}_{+1}\\
+(1-m_z\cos^2\theta_j)e^{i(m_z+1)\phi_j}\eta^{\mu}_{-1}\\
+\frac{1}{\sqrt{2}}\sin2\theta_je^{im_z\phi_j}\eta^{\mu}_{0}],
\end{split}
\end{equation}
where the normalization constant is $A_N=A/N$.
The angles are 
\begin{equation}
\begin{split}
\label{eq:ad5a}
&\cos\phi_j'=\frac{\rho\cos\phi_{\rho}-R\cos\phi_j}{|\vec{\rho}-\vec{r}_j|}\\
&\sin\phi_j'=\frac{\rho\sin\phi_{\rho}-R\sin\phi_j}{|\vec{\rho}-\vec{r}_j|}\\
&\cos\theta_j'=\frac{z}{|\vec{r}-\vec{r}_j|}\\
&\sin\theta_j'=\frac{|\vec{\rho}-\vec{r}_j|}{|\vec{r}-\vec{r}_j|}.
\end{split}
\end{equation}

We are interested in  large distances $z\gg R$, therefore we use an approximation $\frac{1}{|\vec{r}-\vec{r}_j|}\approx \frac{1}{z}$. The field is
\begin{equation}
\label{eq:ad5}
\begin{split}
\bra{0}\hat{A}^{\mu}\ket{m_z}_j=A\frac{\omega_0^2e^2}{3\epsilon_0c^2}\frac{1}{z}e^{ik|\vec{r}-\vec{r_j}|}e^{-i\omega t}\\
[\frac{3}{16\pi}(1+m_z\cos^2\theta'_j)e^{i(m_z-1)\phi'_j}\eta^{\mu}_{+1}\\
+\frac{3}{16\pi}(1-m_z\cos^2\theta'_j)e^{i(m_z+1)\phi'_j}\eta^{\mu}_{-1}\\
+\frac{3}{16\sqrt{2}\pi}\sin2\theta'_je^{im_z\phi'_j}\eta^{\mu}_{0}].
\end{split}
\end{equation}

We also assume that $z\gg R \gg \rho$, $\frac{1}{|\vec{\rho}-\vec{r}_j|}\to\frac{1}{R}$. Then
\begin{equation}
\label{eq:ad6}
\begin{split}
\cos\phi_j'=\frac{1}{R}(\rho \cos\phi_{\rho}-R\cos\phi_j)\\
\sin\phi_j'=\frac{1}{R}(\rho\sin\phi_{\rho}-R\sin\phi_j) 
\end{split}
\end{equation}
and
\begin{equation}
\label{eq:ad7}
\begin{split}
e^{i2\phi_j'}=\frac{1}{R^22}(\rho e^{i\phi_{\rho}}-Re^{i\phi_j})^2\approx e^{i2\phi_j},
\end{split}
\end{equation}
therefore $\phi_j'\approx\phi_j$.

Note that $\theta_k$ is defined as $\theta_k=\arctan(\frac{R}{z})\approx \frac{R}{z}$. Further expanding
\begin{equation}
\label{eq:ad8}
\begin{split}
\cos^2\theta_j'= \frac{z^2}{z^2+\rho^2+R^2-2R\rho\cos(\phi_{\rho}-\phi_j)}&\approx1-\theta_k^2\\
\sin^2\theta_j'=\frac{\rho^2+R^2-2R\rho\cos(\phi_{\rho}-\phi_j)}{z^2+\rho^2+R^2-2R\rho\cos(\phi_{\rho}-\phi_j)}
&\approx\theta_k^2,
\end{split}
\end{equation}
leads to approximately equal $\theta_j'\approx\theta_k$.

The total vector field of the atomic ring array with finite number of sources is
\begin{equation}
\label{eq:d13}
\begin{split}
&A^{\mu} _{m_z}(\vec{r},t)=\frac{A_N}{16\pi}\frac{\omega_0^2e^2}{\epsilon_0c^2}\frac{1}{z}e^{-i\omega t}\\ \sum_{j=0}^{N-1} [
&(1+m_z\cos^2\theta_k)e^{i(m_z-1)\phi_j}\eta^{\mu}_{+1}\\
&+(1-m_z\cos^2\theta_k)e^{i(m_z+1)\phi_j}\eta^{\mu}_{-1}\\
&+\frac{1}{\sqrt{2}}\sin2\theta_ke^{im_z\phi_j}\eta^{\mu}_{0}]e^{il\phi_j}e^{ik|\vec{r}-\vec{r}_j|}.
\end{split}
\end{equation}

The sum over sources $j$ can be evaluated using Jacobi-Anger expansion:
\begin{equation}
\label{eq:d14}
\begin{split}
&\sum_{j=0}^{N-1}e^{i(l+m_z)\phi_j}e^{ik|\vec{r}-\vec{r}_j|}\\
&=e^{ik(z+\frac{R^2}{z})}\sum_{j=0}^{N-1}e^{i(l+m_z)\phi_j}e^{-i\kappa\rho\cos(\phi_{\rho}-\phi_j)}\\
&=Ne^{ik(z+\frac{R^2}{z})}\sum_{j=0}^{N-1}\sum_{n=-\infty}^{n=\infty}(-i)^nJ_{n}(\kappa\rho)e^{in(\phi_{\rho}-\phi_j)}e^{i(l+m_z)\phi_j}.
\end{split}
\end{equation}
The above result corresponds to the $\eta_0^\mu$ component of vector field in Eq.(\ref{eq:d13}). For $+1$ and $-1$ spherical-vector components ($i.e.$, the factors in front of $\eta_{+1}^\mu$  and $\eta_{-1}^\mu$ in Eq.(\ref{eq:d13})), we have  
\begin{equation}
\label{eq:d15}
\begin{split}
Ne^{ik(z+\frac{R^2}{z})}\sum_{j=0}^{N-1}\sum_{n=-\infty}^{n=\infty}(-i)^nJ_{n}(\kappa\rho)e^{in(\phi_{\rho}-\phi_j)}e^{i(l+m_z-1)\phi_j},
\end{split}
\end{equation}
and
\begin{equation}
\label{eq:d16}
\begin{split}
Ne^{ik(z+\frac{R^2}{z})}\sum_{j=0}^{N-1}\sum_{n=-\infty}^{n=\infty}(-i)^nJ_{n}(\kappa\rho)e^{in(\phi_{\rho}-\phi_j)}e^{i(l+m_z+1)\phi_j}.
\end{split}
\end{equation}

Summation over $j$ selects certain orders of Bessel vortices $n$, while other order terms are canceled, as was demonstrated for scalar fields in \cite{Wang_2022}. After summation of radiation from individual atoms, the final result for the field is:

\begin{equation}
\label{eq:d13a}
\begin{split}
&A^{\mu}_{m_z}(\vec{r},t)=\frac{A}{16\pi}\frac{\omega_0^2e^2}{\epsilon_0c^2}\frac{1}{z}e^{-i\omega t}e^{ik(z+\frac{R^2}{z})} \\
&[
(1+m_z\cos^2\theta_k)C_{+1}\eta^{\mu}_{+1}
+(1-m_z\cos^2\theta_k)C_{-1}\eta^{\mu}_{-1}\\
&+\frac{1}{\sqrt{2}}\sin2\theta_k C_0\eta^{\mu}_{0}],
\end{split}
\end{equation}
where the coefficient for $\eta_0^\mu$ field component  is
\begin{equation}
\label{eq:d17}
\begin{split}
C_0=\sum_{m=-\infty}^{m=\infty}[(-i)^nJ_{n}(\kappa\rho)e^{in\phi_{\rho})}]_{n=l+m_z+mN},
\end{split}
\end{equation}
and for $+1$ and $-1$ field components we obtain, 
\begin{equation}
\label{eq:d18}
\begin{split}
C_{\pm1}=\sum_{m=-\infty}^{m=\infty}[(-i)^nJ_{n}(\kappa\rho)e^{in\phi_{\rho}}]_{n=l+m_z\mp1+mN},
\end{split}
\end{equation}
where $m$ is an integer that also serves as a summation index: $m=0,\pm 1,\pm 2,...$.

The above equations Eqs.(\ref{eq:d13a}-\ref{eq:d18}) represent a result for the total vector potential of a phased atomic array with $N$ atoms and a phase parameter $l$ expressed in terms of an infinite series of Bessel vortices.
The quantity $m=0$ corresponds to the leading order terms $n=l+m_z\pm 1$ and $n=l+m_z$ of each of the three spherical components of the vector vortex field. In general, we have an infinite series of Bessel vortices contributing. But a closer look reveals a crucial role of the number of emitters $N$: the next non-vanishing order corresponds to addition/substraction of an integer number of $N$, so the larger is $N$, the wider is the ``gap" between the orders of contributing Bessel vortices. 

The result also sets a requirement on a minimal number of atoms necessary to form a vortex. To generate a vortex of order $l$ for each field component, we find the necessary conditions $N\geq2(l+m_z)-1$ for the positive-helicity component; $N\geq2(l+m_z)+3$ for the negative-helicity component and $N\geq2(l+m_z)+1$ for the zeroth (or longitudinal) component are satisfied. The strongest of above conditions is $N\geq2(l+m_z)+3$. If it is satisfied, then all three components of the field have vortex structure ($i.e.$, show a phase singularity). When $m_z=-1$, the condition becomes $N\geq 2l+1$ which is the condition for a scalar vortex field \cite{Wang_2022}. However, for $m_z=+1$, the conditions are different. The difference is due to the different azimuthal variation of the phase  $e^{i(l+m_z-1)\phi_{\rho}}$, $e^{i(l+m_z+1)\phi_{\rho}}$ and $e^{i(l+m_z)\phi_{\rho}}$  in the three spherical components of the vector vortex field. 

If we only keep the leading-order terms in Eqs.(\ref{eq:d17}-\ref{eq:d18}), the vortex field exactly matches the continuous ($N\to\infty$) limit given by Eq.(\ref{eq:d12}).

Let us discuss the role of spin in our approach. For circularly polarized Bessel beams carrying OAM, the helicity $\Lambda=\pm 1$ plays the role of spin angular momentum projection \cite{PhysRevA.88.033841} in the limit of small opening angles $\theta_k$. In our case, the rays of light forming an optical vortex are generated by the atoms from an array. As required by conservation of angular momentum, the corresponding atomic states $m_z=\pm 1$ emit right- or left-circularly polarized light, if the quantization axis is along the ray's direction. 
Atom's magnetic quantum number $m_z$ in our theory plays the role of the spin angular momentum projection of the generated vortex field, if taken in a paraxial approximation $\theta_k\ll 1 $. The quantity $l+m_z$ therefore is  similar to the total angular momentum projection of the vortex field, $m_{\gamma}=l+\Lambda$ for Bessel beam with circular polarization \cite{PhysRevA.88.033841}. 


\section{Polarization Parameters of Atomic-Array Radiation}

Presence of a longitudinal electric field component in the vector vortices requires definition of 3x3 spin density matrix described by nine independent parameters (including field's intensity). One possibility is to use expansion coefficients of nine Gell-Mann matrices \cite{PhysRevA.90.023809,Eismann:2021te}.
It is also possible to describe the electromagnetic field in terms of state multipoles.  Using such an approach, it was shown in Ref.\cite{Afanasev_2020} that polarization parameters of twisted photons determine a spin density matrix of optically polarized atoms.

Let us briefly describe the formalism. An arbitrary (pure) state of spin-1 particle $|\chi_1 \rangle$, or a 3-dimensional vector field, can be expanded in terms of basis kets $|\chi_{1i}\rangle$ $(i=x,y,z)$ in a Cartesian basis, 
$|\chi_1\rangle= \sum_i a_i |\chi_{1i}\rangle$, as well as in a spherical basis $$|\chi_1 \rangle=a^+|\chi_{1+} \rangle+a^0|\chi_{10} \rangle+a^-|\chi_{1-} \rangle,$$  where $a_i$ and $a^{\pm}$ are complex amplitudes related by  $$ a^\pm=\frac{(\mp a_x+i a_y)}{\sqrt{2}}, \, a_0=a_z, \, \sum_i |a_i|^2=1.$$
  
The spin density matrix is  $\rho_{ij}=a_i a_j^*$, and we will use parameterization and normalization conventions from Ref.\cite{Afanasev_2020,Ohlsen_1972}:
\begin{eqnarray}
\rho_{ij}=&&\frac{1}{3}\Big\{ I+\frac{3}{2}(p_x \mathcal P_x+p_y \mathcal P_y+p_z \mathcal P_z)+ \nonumber\\
&&\frac{2}{3}(p_{xy} \mathcal P_{xy}+p_{yz} \mathcal P_{yz}+p_{xz} \mathcal P_{xz})+\\
&&\frac{1}{6}(p_{xx}-p_{yy})(\mathcal P_{xx}-\mathcal P_{yy})+\frac{1}{2}p_{zz}\mathcal P_{zz}\Big\}_{ij}, \nonumber
\end{eqnarray}
where $\mathcal P_i,\, \mathcal P_{ij}$ are the operators of spin and quadrupole moment,  and $p_i,\, p_{ij}$ are corresponding vector and quadrupole polarizations that can be expressed in terms of the above amplitudes $a_i$ and $a^{\pm,0}$,
\begin{equation}
\label{eq:poldefs}
p_i=i \sum_{jk} \epsilon_{ijk}a_j a_k^*; \, p_{ik}=-\frac{3}{2} (a_i a_k^*+a_k a_i^*-\frac{2}{3}\delta_{ik}).
\end{equation}
In particular,
\begin{align}
\label{eq:qpol}
&p_{xx}-p_{yy}=3 (a^+a^{-*}+a^-a^{+*});\nonumber\\ &p_{zz}=|a^+|^2+|a^-|^2-2|a^0|^2 ; \nonumber\\
&p_{xy}=i\frac{3}{2}(a^+ a^{-*}-a^-a^{+*}) \\
&p_{xz}=\frac{3}{2\sqrt{2}}(a^+a^{0*}+a^0a^{+*}-a^-a^{0*}-a^{0}a^{-*}) \nonumber \\
&p_{yz}=i\frac{3}{2\sqrt{2}}(a^+a^{0*}-a^0a^{+*}+a^-a^{0*}-a^{0}a^{-*}). \nonumber
\end{align}
The components of polarization have the following bounds, $-3\leq p_{xx}-p_{yy} \leq 3$, $-2\leq p_{ii} \leq 1$, $-\frac{3}{2}\leq p_{ij}\leq \frac{3}{2}$ ($i\neq j$), and $-1\leq p_i\leq1$. These quantities are commonly referred to in atomic and nuclear physics as orientation $p_i$ and alignment $p_{ik}$. For a comprehensive formalism of spin-1 particle polarization, see also Ref.~\cite{varshalovich1988quantum}, noting that the definition of $p_{nm}$ used here have an extra factor of 3. Another convention for the description of optical polarisation in 3D fields uses an expansion in terms of Gell-Mann matrices \cite{carozzi2000}, and is equivalent to the approach presented here.

Let us derive expressions for the vector polarization $p_z$ (or orientation) and quadrupole polarization $p_{zz}$ (or alignment) of the vector vortex field generated by the phased ring array of atoms. The simplest expressions are obtained for $N\to\infty$  with the electric field from the vector potential Eq.(\ref{eq:d12}):
\begin{equation}
\label{eq:bd1}
    \begin{split}
       E^{\mu}_{m_z}=\frac{\partial }{\partial t} A^{\mu} _{m_z}=-i\omega A^{\mu} _{m_z}=-i\omega Ae^{-i(\omega t-k_zz)}\sqrt{\frac{\kappa}{2\pi}}\\(-\frac{ i}{2}e^{i(l+m_z-1)\phi_{\rho}}J_{l+m_z-1}(\kappa\rho)(1+m_z\cos^2\theta_k)\eta_{1}^{\mu}\\+\frac{ i}{2}e^{i(l+m_z+1)\phi_{\rho}}J_{l+m_z+1}(\kappa\rho)(1-m_z\cos^2\theta_k)\eta_{-1}^{\mu}\\+\frac{m_z }{2\sqrt{2}}e^{i(l+m_z)\phi_{\rho}}J_{l+m_z}(\kappa\rho)\sin2\theta_k\eta_{0}^{\mu}).\\
    \end{split}
\end{equation}

To evaluate orientation and alignment parameters, $p_z$ and $p_{zz}$, we use the definitions from Eqs.(\ref{eq:poldefs},\ref{eq:qpol}) and the electric field components from the above equation, 
\begin{equation}
\label{eq:d24}
\begin{split}
p_z=\frac{|E^{\mu}_{+1}|^2-|E^{\mu}_{-1}|^2}{|E^{\mu}_{+1}|^2+|E^{\mu}_{-1}|^2+|E^{\mu}_0|^2}.
\end{split}
\end{equation}
\begin{equation}
\label{eq:zd2}
\begin{split}
p_{zz}=\frac{|E^{\mu}_{+1}|^2+|E^{\mu}_{-1}|^2-2|E^{\mu}_{0}|^2}{|E^{\mu}_{+1}|^2+|E^{\mu}_{-1}|^2+|E^{\mu}_0|^2}.
\end{split}
\end{equation}

We are interested in particular in far-field polarization properties of array's radiation in the vicinity of phase singularity, where $\theta_k\approx\frac{R}{z}\ll 1$, therefore Taylor expansion in $\theta_k$  is justified.  Keeping only the leading terms in $\theta_k$, we obtain for the 
positive-helicity component ($\eta_1$),
\begin{equation}
\label{eq:d21}
\begin{split}
&(1+m_z\cos^2\theta_k)(-i)^{-1}J_{l+m_z-1}(\kappa\rho)\approx\\
&(1+m_z-m_z\theta_k^2))(-i)^{-1}\frac{1}{(l+m_z-1)!}(\frac{x\theta_k}{2})^{|l+m_z-1|},
\end{split}
\end{equation}
where $x=k \rho=\frac{2\pi}{\lambda}\rho$.
For the negative-helicity component ($\eta_{-1}$) we have,
\begin{equation}
\label{eq:d22}
\begin{split}
&(1-m_z\cos^2\theta_k)(-i)J_{l+m_z+1}(\kappa\rho)\approx\\
&(1-m_z+m_z\theta_k^2)(-i)\frac{1}{(l+m_z+1)!}(\frac{x\theta_k}{2})^{|l+m_z+1|},
\end{split}
\end{equation}
and for the longitudinal field component ($\eta_0$),
\begin{equation}
\label{eq:d23}
\begin{split}
&\frac{1}{\sqrt{2}}\sin2\theta_kJ_{l+m_z}(\kappa\rho)\approx\\
&\sqrt{2}\theta_k\frac{1}{(l+m_z)!}(\frac{x\theta_k}{2})^{|l+m_z|},
\end{split}
\end{equation}
where we left out a common factor that can be restored from Eq.(\ref{eq:bd1}).

Substituting $m_z=+1$ in the above expressions,  we obtain
\begin{equation}
\label{eq:d27}
\begin{split}
|E^{\mu}_{-1}|^2&\sim \theta_k^{2(l+4)}\frac{1}{(l+2)!^2}(\frac{x}{2})^{2|l+2|}\\
|E^{\mu}_{+1}|^2&\sim \frac{4}{l!^2}(\frac{x}{2})^{2l}\theta_k^{2|l|}\\
|E^{\mu}_{0}|^2&\sim \frac{2\theta_k^{2(l+2)}}{(l+1)!^2}(\frac{x}{2})^{2|l+1|},
\end{split}
\end{equation}
where a common factor is omitted since it does not affect polarization parameters.
The term $|E^{\mu}_{+1}|^2$ contains the lowest order of $\theta_k^{2l}$. Therefore in the leading order of $\theta_k$-expansion 
\begin{equation}
\label{eq:d28}
\begin{split}
p_z\approx\frac{|E^{\mu}_{+1}|^2}{|E^{\mu}_{+1}|^2}=1\\
p_{zz}\approx\frac{|E^{\mu}_{+1}|^2}{|E^{\mu}_{+1}|^2}=1.
\end{split}
\end{equation}
Therefore both $p_z$ (orientation) and $p_{zz}$ (alignment) are equal to unity in the leading order of $\theta_k$, $i.e.$, in the far field case, provided that $l$ and $m_z$ parameters have the $same$ sign. The same result, $p_z=p_{zz}=1$ is obtained for $l$=0.

When $m_z=-1$, we have for the field components
\begin{equation}
\label{eq:d29}
\begin{split}
|E^{\mu}_{-1}|^2&\sim \frac{4\theta_k^{2l}}{l!^2}(\frac{x}{2})^{2|l|}\\
|E^{\mu}_{+1}|^2&\sim \frac{\theta_k^{2l}}{(l-2)!^2}(\frac{x}{2})^{2|l-2|}\\
|E^{\mu}_{0}|^2&\sim \frac{2\theta_k^{2l}}{(l-1)!^2}(\frac{x}{2})^{2|l-1|},
\end{split}
\end{equation}
where $|E^{\mu}_{-1}|^2$ and $|E^{\mu}_{0}|^2$ have the same orders $\theta_k^{2(l+m_z)+2}$ and $|E^{\mu}_{+1}|^2$ remains the order $\theta_k^{2(l+m_z)+2}$. 

\begin{equation}
\label{eq:d31}
\begin{split}
p_z=\frac{\frac{1}{(l-2)!^2}(\frac{x}{2})^{2|l-2|}-\frac{4}{l!^2}(\frac{x}{2})^{2|l|}}{\frac{1}{(l-2)!^2}(\frac{x}{2})^{2|l-2|}+\frac{4}{l!^2}(\frac{x}{2})^{2|l|}+\frac{2}{(l-1)!^2}(\frac{x}{2})^{2|l-1|}}.
\end{split}
\end{equation}

\begin{equation}
\label{eq:d32}
\begin{split}
p_{zz}=\frac{\frac{1}{(l-2)!^2}(\frac{x}{2})^{2|l-2|}+\frac{4}{l!^2}(\frac{x}{2})^{2|l|}-\frac{4}{(l-1)!^2}(\frac{x}{2})^{2|l-1|}}{\frac{1}{(l-2)!^2}(\frac{x}{2})^{2|l-2|}+\frac{4}{l!^2}(\frac{x}{2})^{2|l|}+\frac{2}{(l-1)!^2}(\frac{x}{2})^{2|l-1|}}.
\end{split}
\end{equation}

Specifically for $l=0$, $m_z=-1$, 
\begin{equation}
\label{eq:ad31}
\begin{split}
p_z&=-1,\\
p_{zz}&=1.
\end{split}
\end{equation}

For $l=1$, $m_z=-1$,
\begin{equation}
\label{eq:bd31}
\begin{split}
p_z&=\frac{-(\frac{x}{2})^{2}}{(\frac{x}{2})^{2}+0.5},\\
p_{zz}&=\frac{(\frac{x}{2})^{2}-1}{(\frac{x}{2})^{2}+0.5}.
\end{split}
\end{equation}

For $l=2$, $m_z=-1$,
\begin{equation}
\label{eq:cd31}
\begin{split}
p_z&=\frac{1-(\frac{x}{2})^{2}}{1+(\frac{x}{2})^{2}},\\
p_{zz}&=\frac{1+(\frac{x}{2})^{4}-4(\frac{x}{2})^{2}}{1+(\frac{x}{2})^{4}+2(\frac{x}{2})^{2}}.
\end{split}
\end{equation}

The derived expressions demonstrate that for {\it opposite-sign} values of $l$ and $m_z$, in the vicinity of phase singularity, the polarization parameters depend only on $l$ and the transverse position with respect to $z$-axis, but are independent on $z$, $i.e.$ on a longitudinal distance to the source array. For the optical vortex beams, a similar effect  was demonstrated to follow from Maxwell's equations \cite{Afanasev2022non}.

\section{Numerical Results and Discussion}

Let us supplement above results with numerical examples. Assume that the wavelength is $\lambda$=1.0$\mu$m and the atoms are prepared in the $m_z=-1$ states. The atomic ring array has a radius of $R=1$mm. We evaluate the energy flux density, $$f(\vec\rho,z)=\cos(\theta_k)(|E|^2+|B|^2)/4,$$ and polarization parameters $p_z$ and $p_{zz}$ for several  distances $z=z_R,1.5z_R,2z_R$ from the array, where $z_R=\frac{\pi w_0^2}{\lambda}=3.14$m is the Rayleigh range  and $w_0=R=1000\lambda$ is beam's waist equal to the array radius. For each value of phase parameter $l=1,2,3$, the results are analyzed for different numbers of radiating atoms $N=3,6,12$. 

\begin{figure*}[h]
\centering
\includegraphics[width=0.45\textwidth]{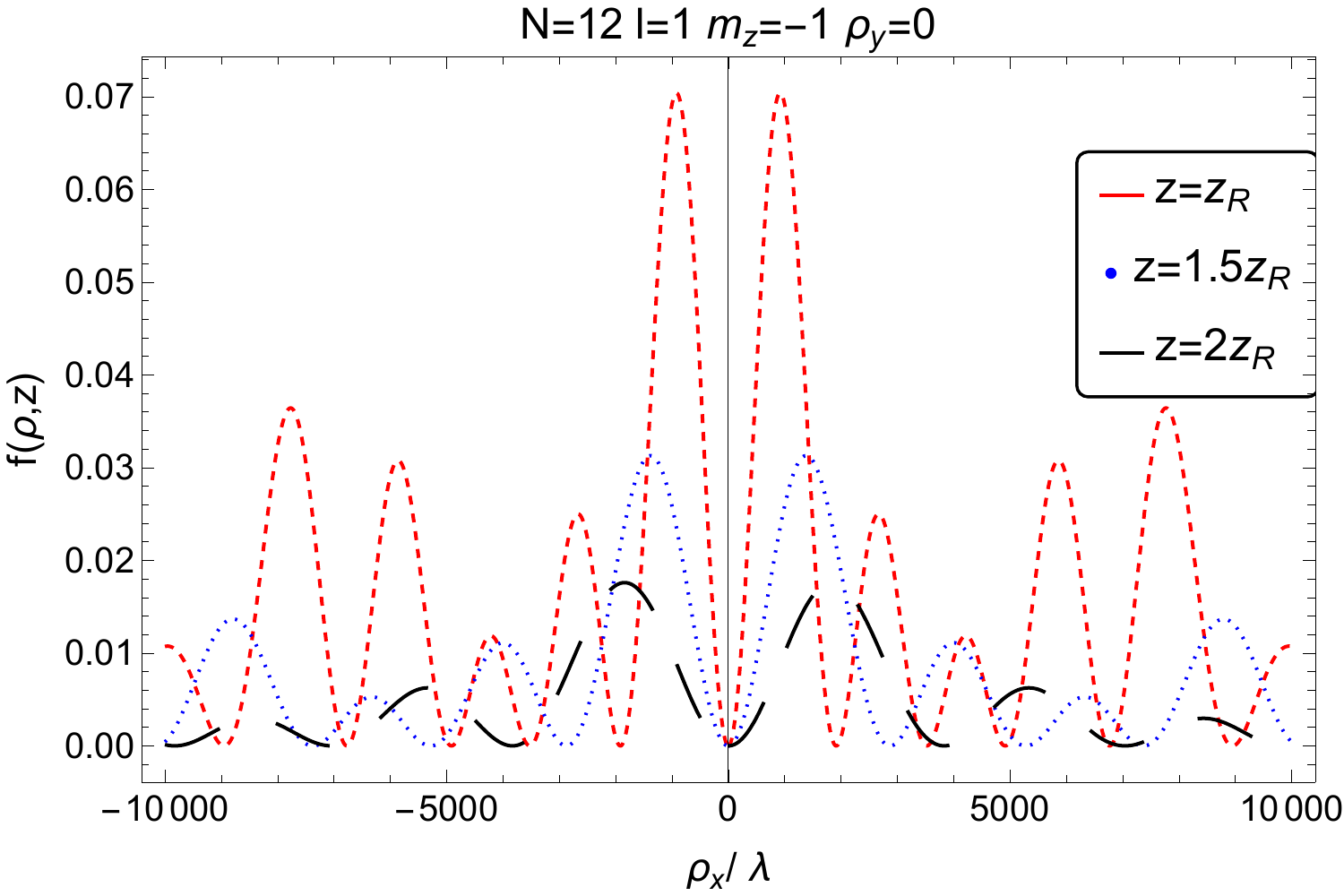}
\includegraphics[width=0.45\textwidth]{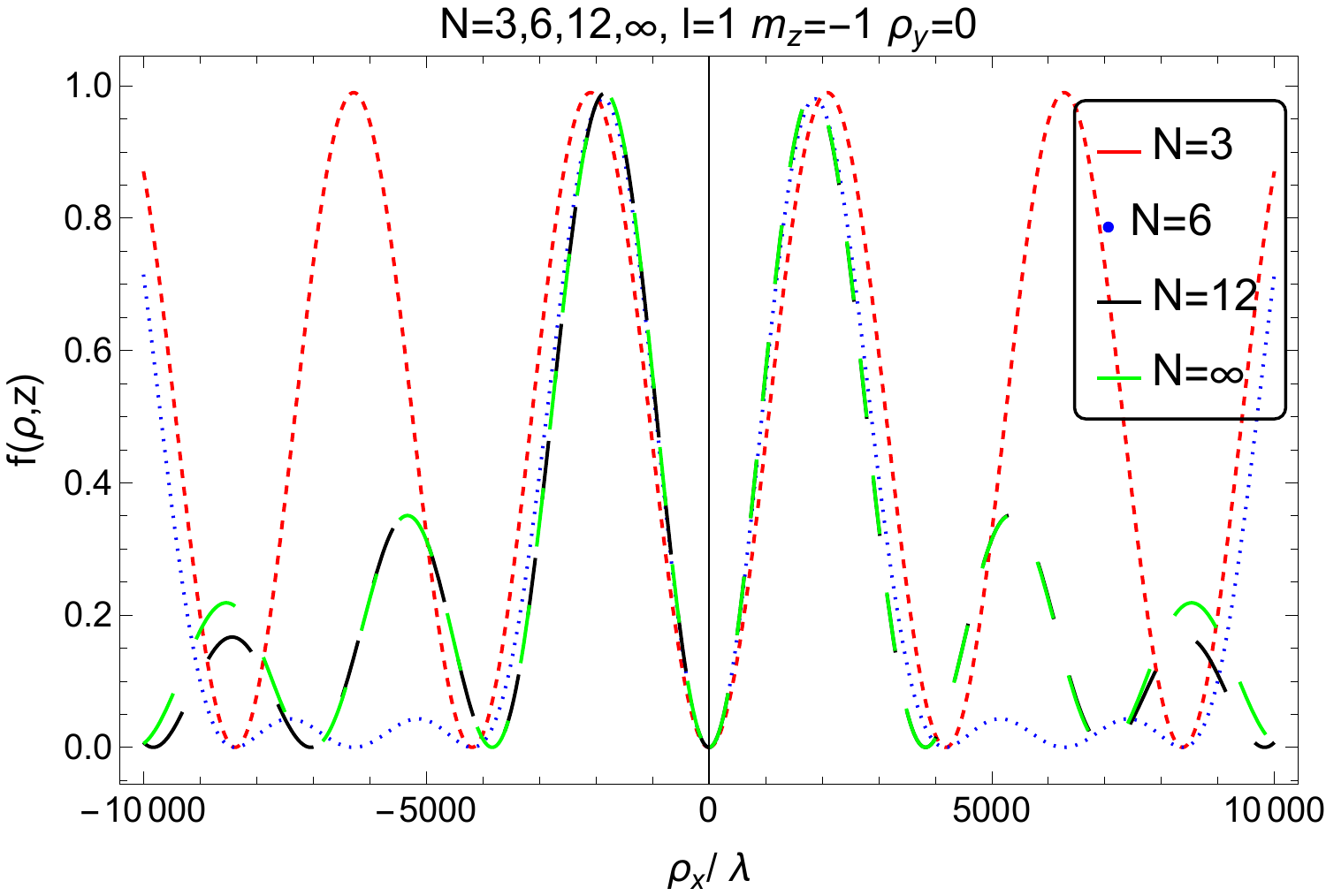}
\caption{Left plot: Flux density  (arbitrary units) against $\rho_x$ on the y-axis when $N=12$, $l=1$ and $m_z=-1$ for three different locations of projection plane along z-axis $z=z_R$ (red dotted), $z=1.5 z_R$ (blue dotted) and $z=2z_R$ (black dashed). Right plot: same as the left plot, but for different number of sources $N=3,6,12,\infty$ ($z=2z_R$)}
\label{fig:a5}
\end{figure*}

\begin{figure*}[htbp]
\centering
\includegraphics[width=0.32\textwidth]{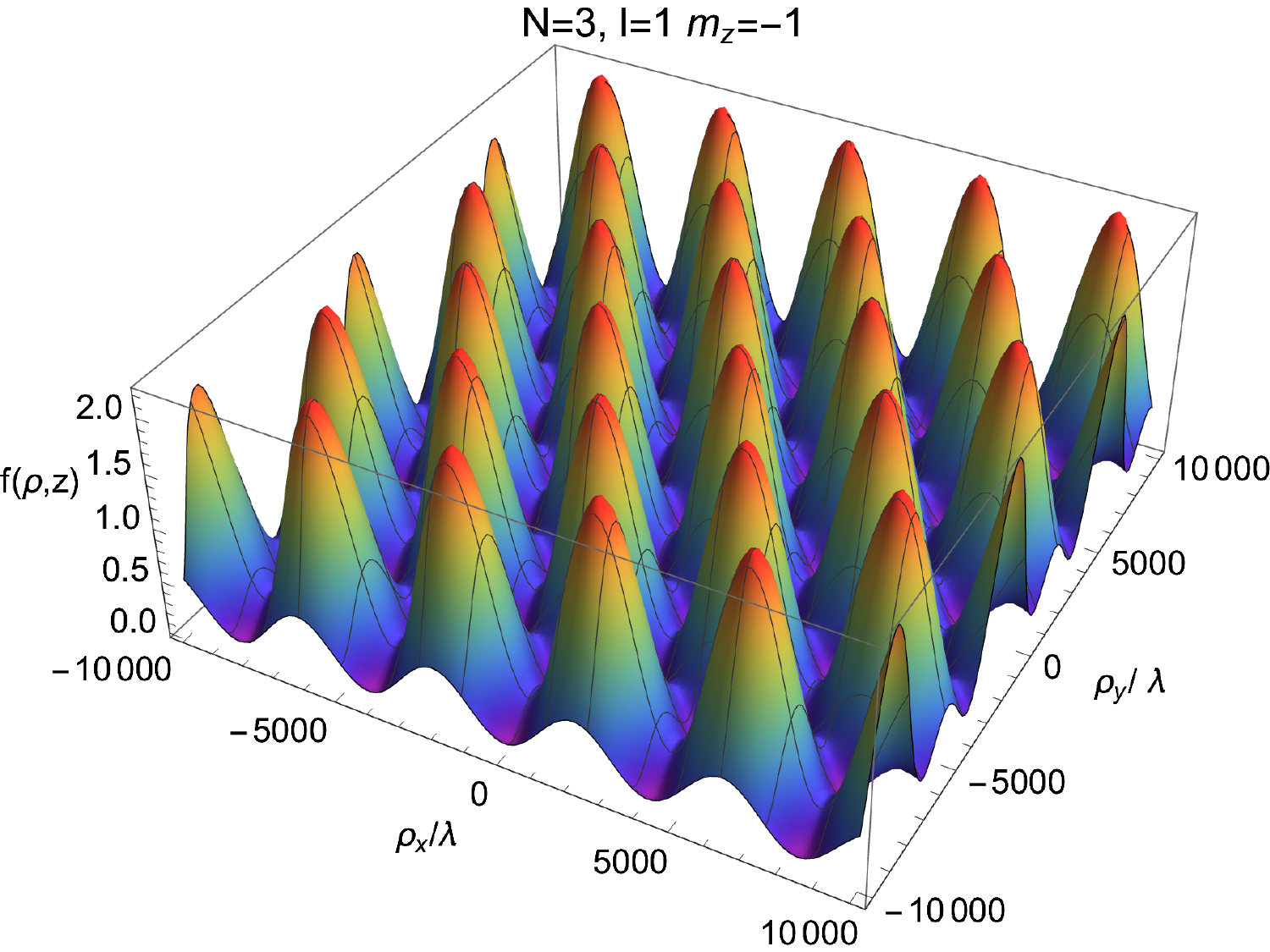}
\includegraphics[width=0.32\textwidth]{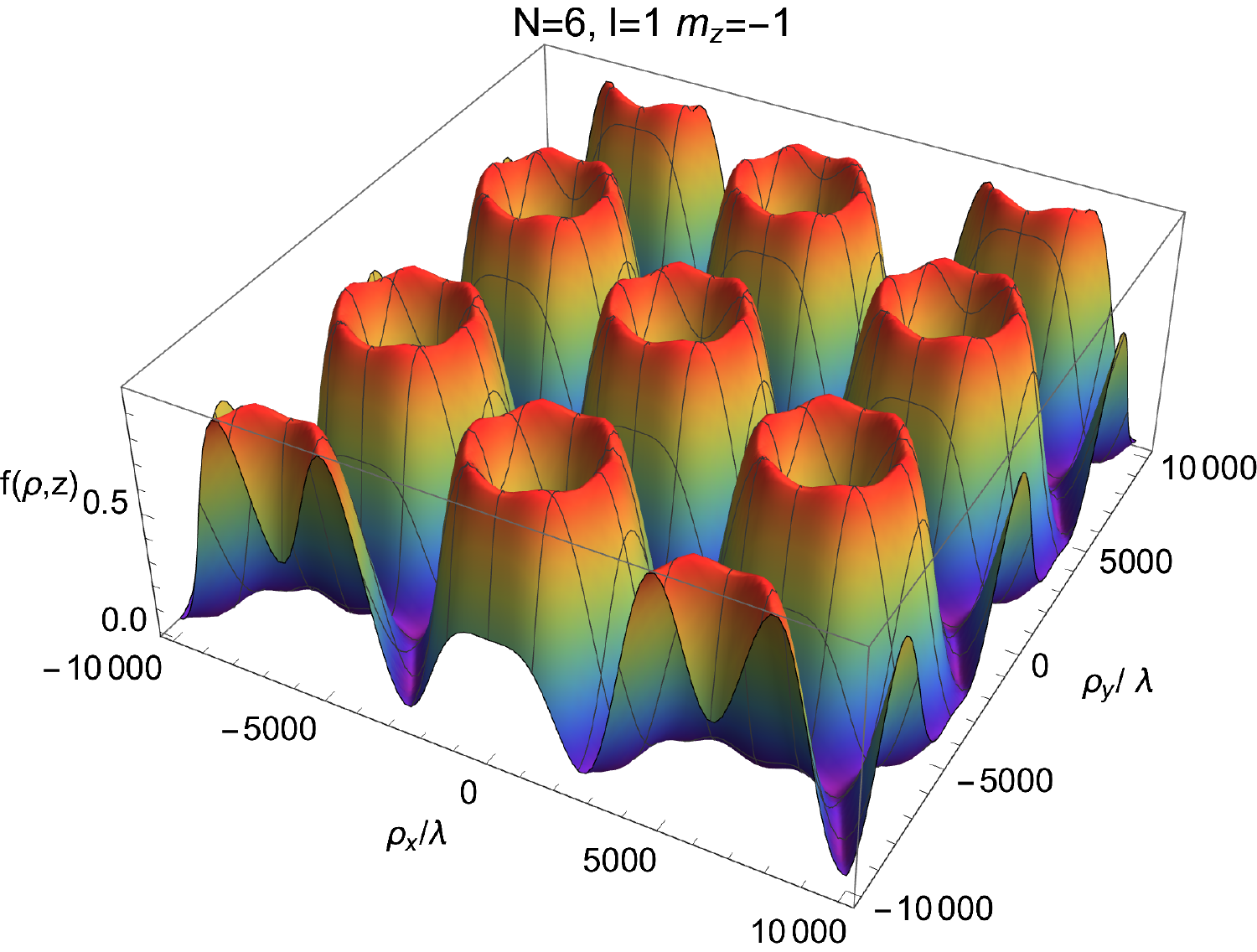}
\includegraphics[width=0.32\textwidth]{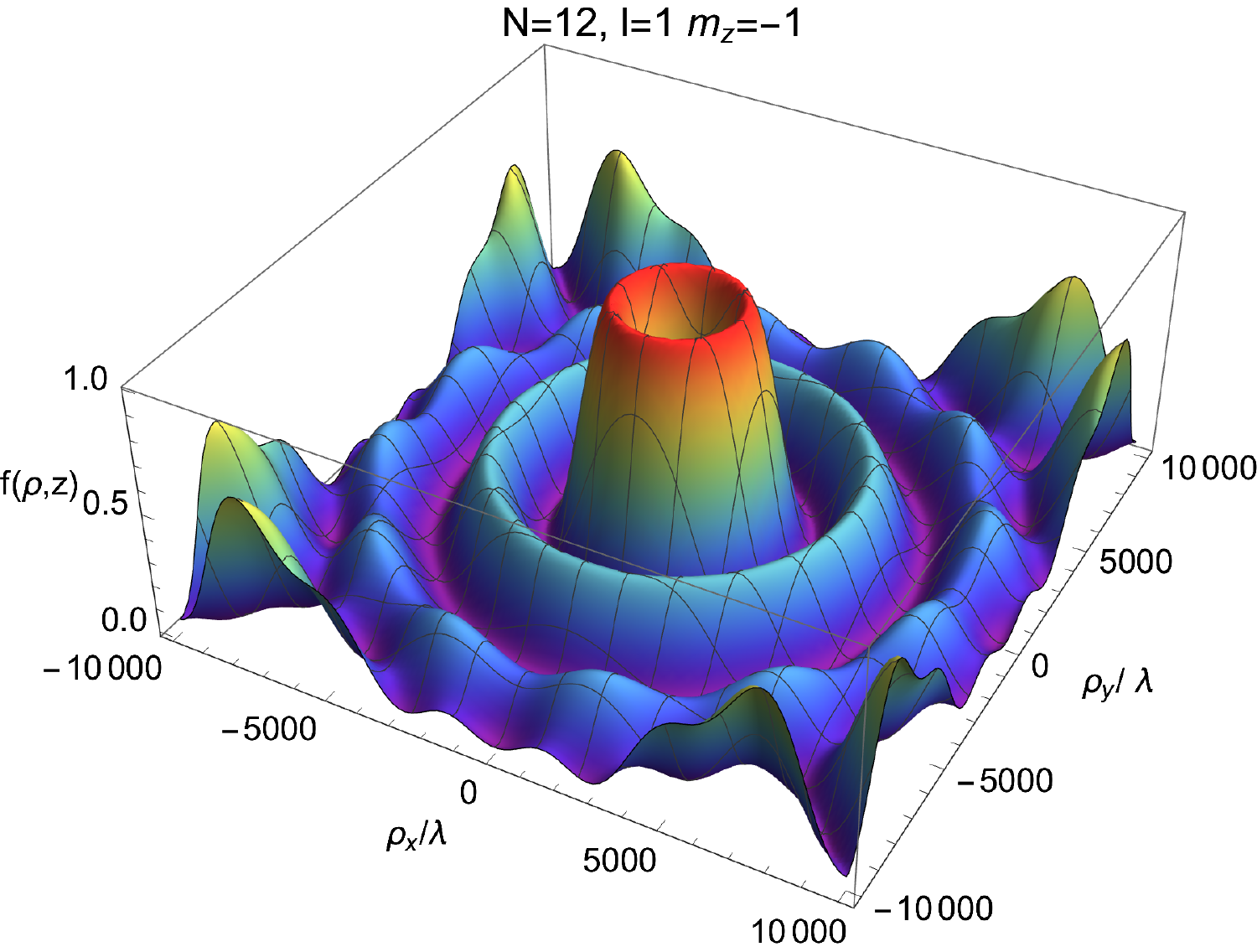}
\caption{Flux density  plotted against transverse position $\rho$  for different numbers of sources $N=$3 (left plot), 6 (middle plot) and 12 (right plot). Here, $l=1$ $m_z=-1$ and $z=2z_R$.}
\label{fig:c1}
\end{figure*}

\begin{figure*}[htbp]
\centering
\includegraphics[width=0.45\textwidth]{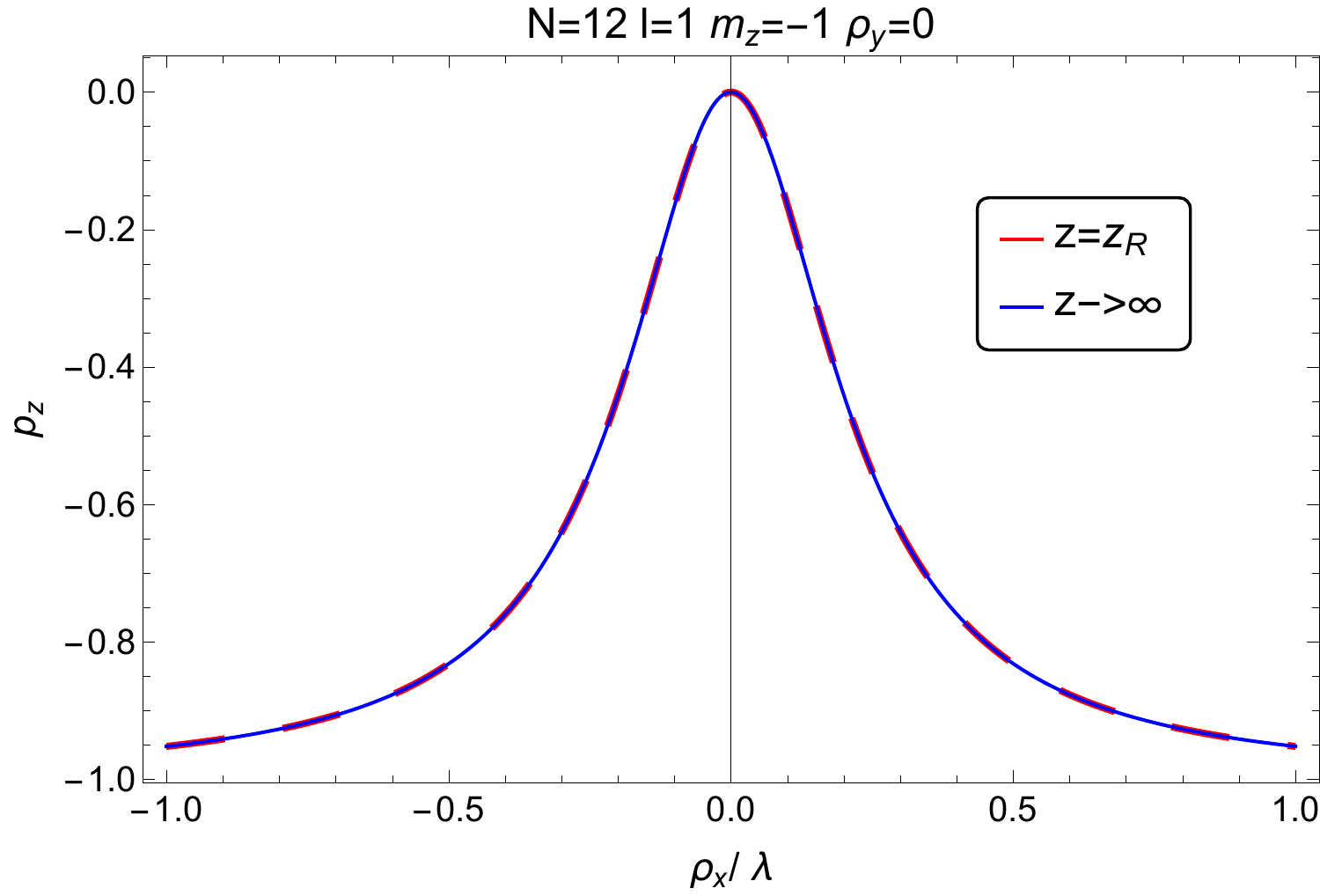}
\includegraphics[width=0.45\textwidth]{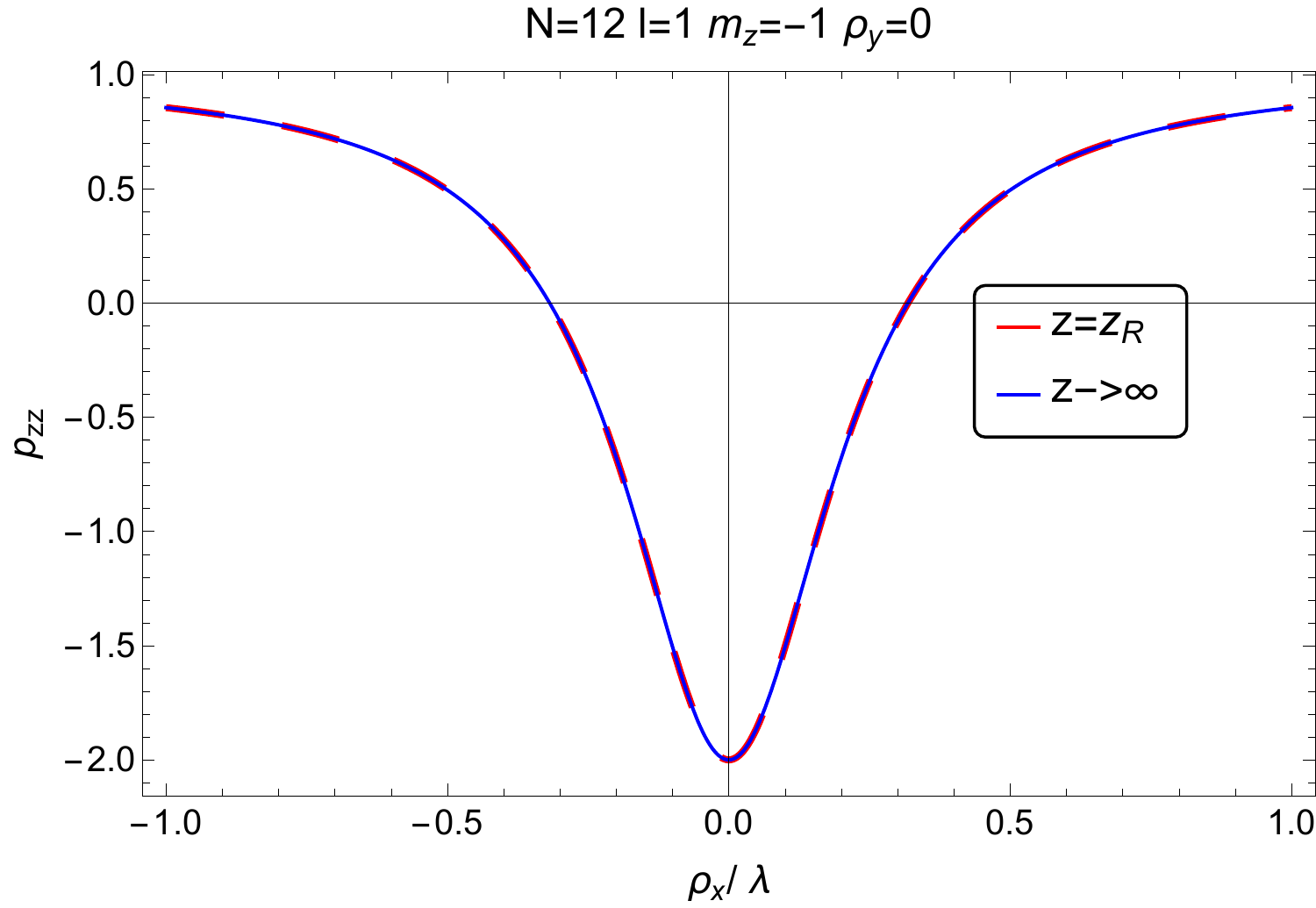}
\caption{Polarization parameter $p_z$ (left plot) and $p_{zz}$ (right plot) against $\rho_x$ on the y-axis in the vicinity of the vortex center ($N=12,l=1,m_z=-1$). The plots at different propagation distances $z$ coincide.}
\label{fig:a2}
\end{figure*}

As shown in Fig.\ref{fig:a5} and Fig.\ref{fig:c1}, the peaks of light energy flux are located in a transverse plane in the range of a few waist radii for $z<2z_R$, that is in 1,000s of wavelengths. We can also see in Fig.\ref{fig:c1} that increasing the number of emitters $N$ results in transition from a lattice of peaks to a ring-like structure (described by three Bessel functions of different orders in the limit $N\to\infty$). For the scalar fields, this transition was previously demonstrated in Ref.\cite{Wang_2022}.  
In contrast to spatial inhomogeneity of the flux, spatial scales for the change of polarizations $p_{z}$ and $p_{zz}$ are about one wavelength around the phase singularity, Fig.\ref{fig:a2}, and, quite remarkably, are independent of propagation distance $z$.

\begin{figure*}[htbp]
\centering
\includegraphics[width=0.45\textwidth]{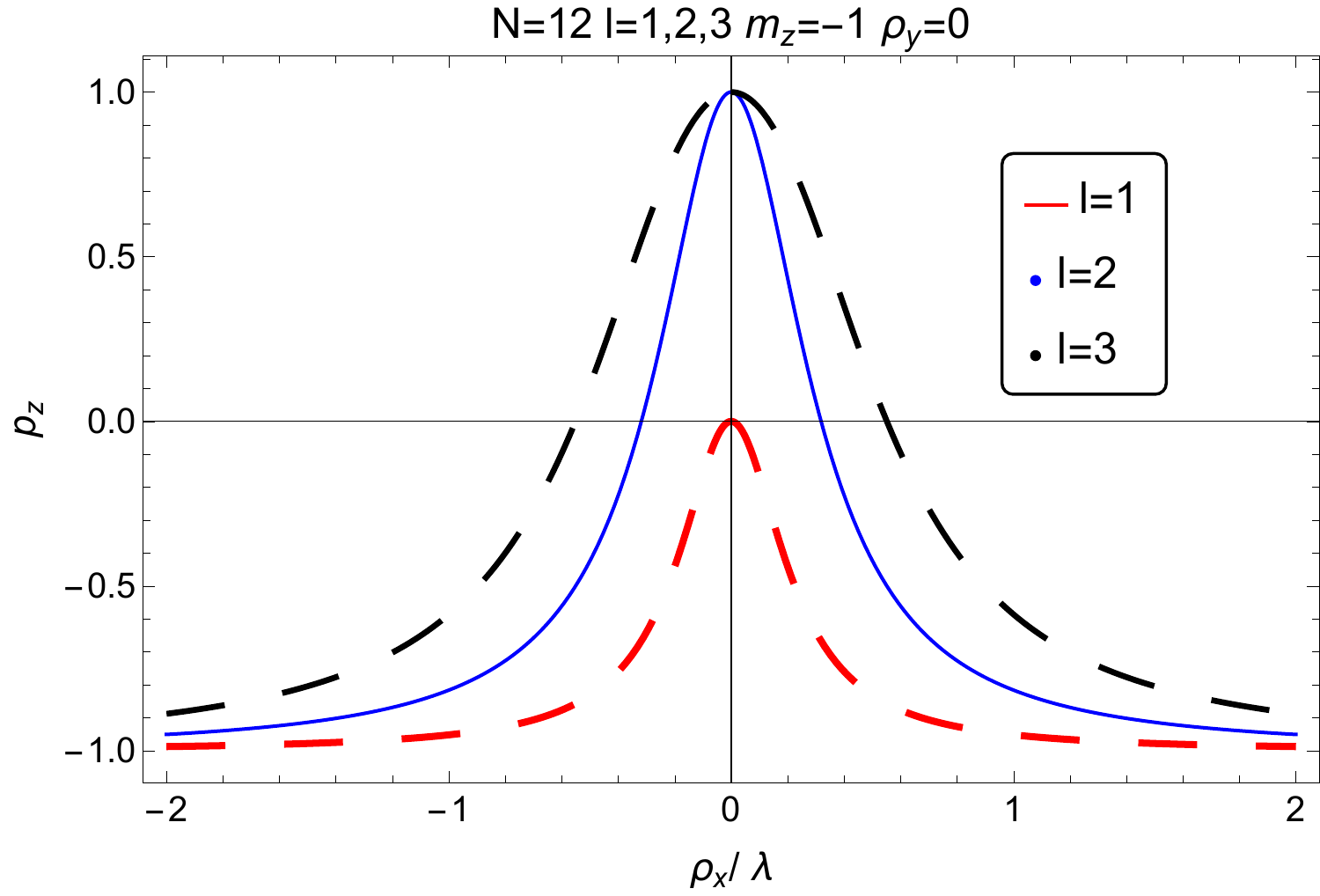}
\includegraphics[width=0.45\textwidth]{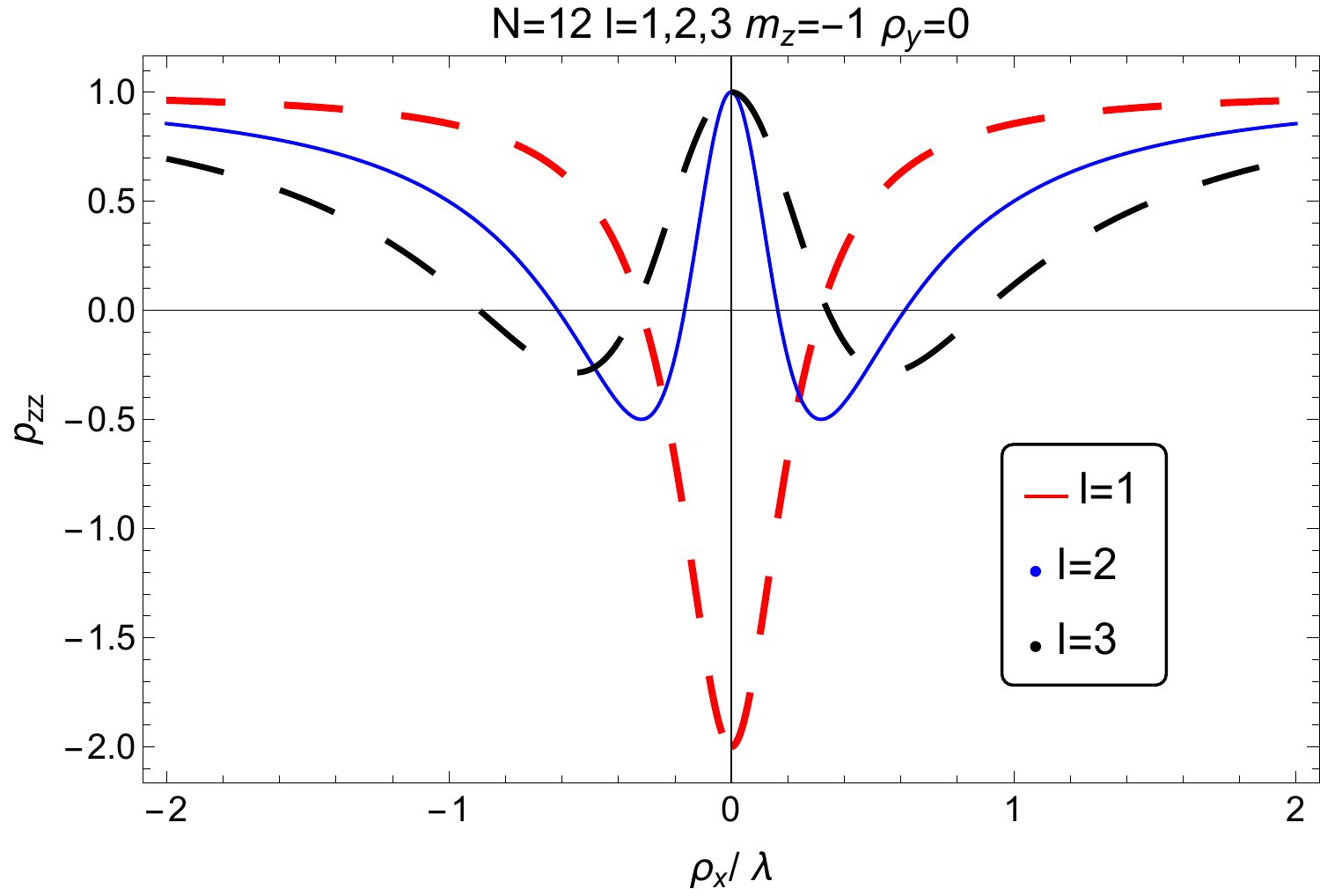}
\caption{Polarization parameter $p_{z}$ (left plot) and $p_{zz}$ (right plot) as a function of radial position $\rho$ for varied phase parameters  $l=1,2,3$ ($m_z=-1$, $z=z_R$ and $N=12$).}
\label{fig:d5}
\end{figure*}

As shown in Fig.\ref{fig:d5}, position-dependence of the polarization parameters near the vortex center is affected by the phase parameter $l$ that for large $N$ may be associated with a topological charge of the electromagnetic vortex. Note from the same Figure that the vector polarization $p_z$ turns to zero in the vortex center if $l$ and $m_z$ are equal in magnitude but opposite in sign. 

\begin{figure*}[htbp]
\centering
\includegraphics[width=0.45\textwidth]{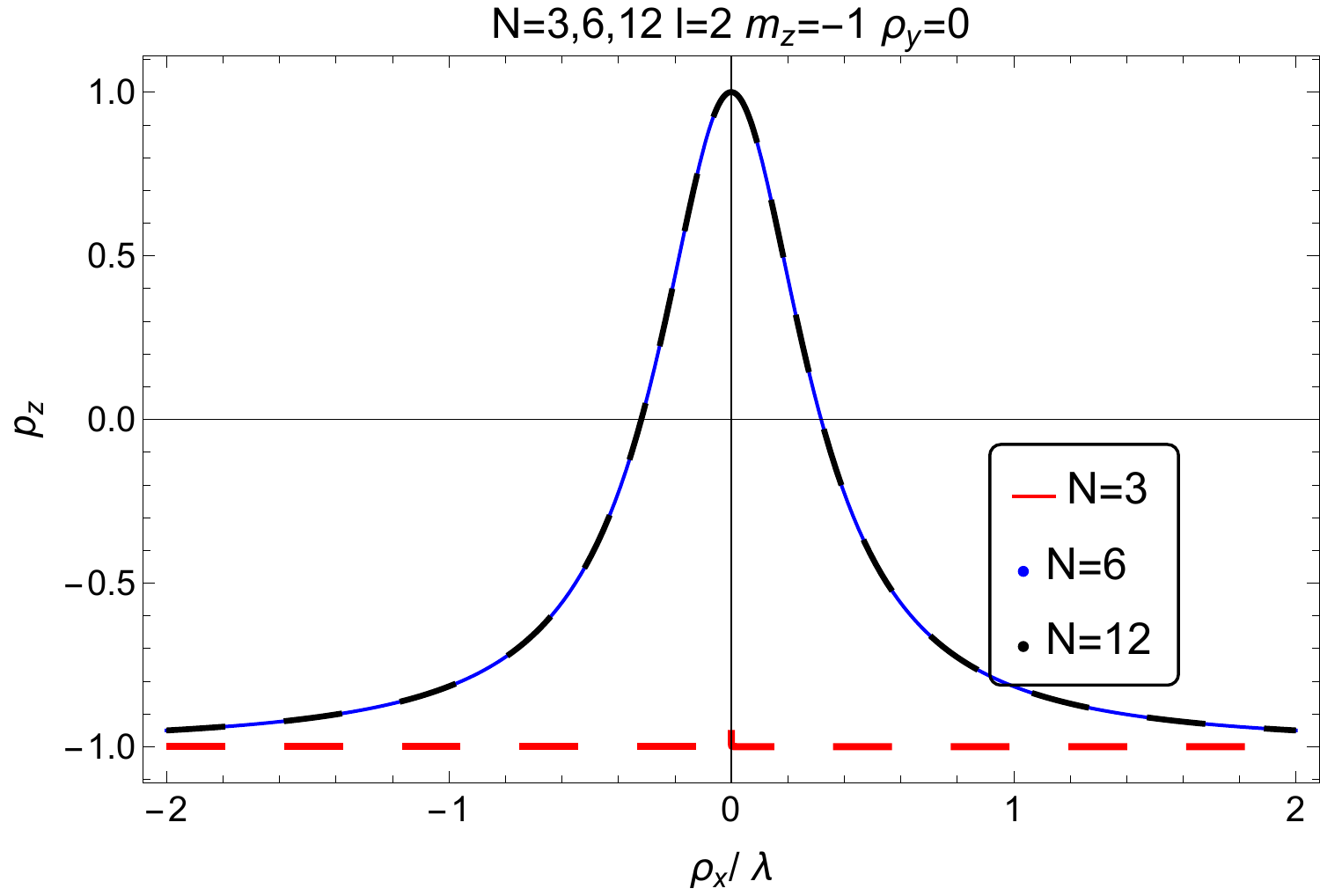}
\includegraphics[width=0.45\textwidth]{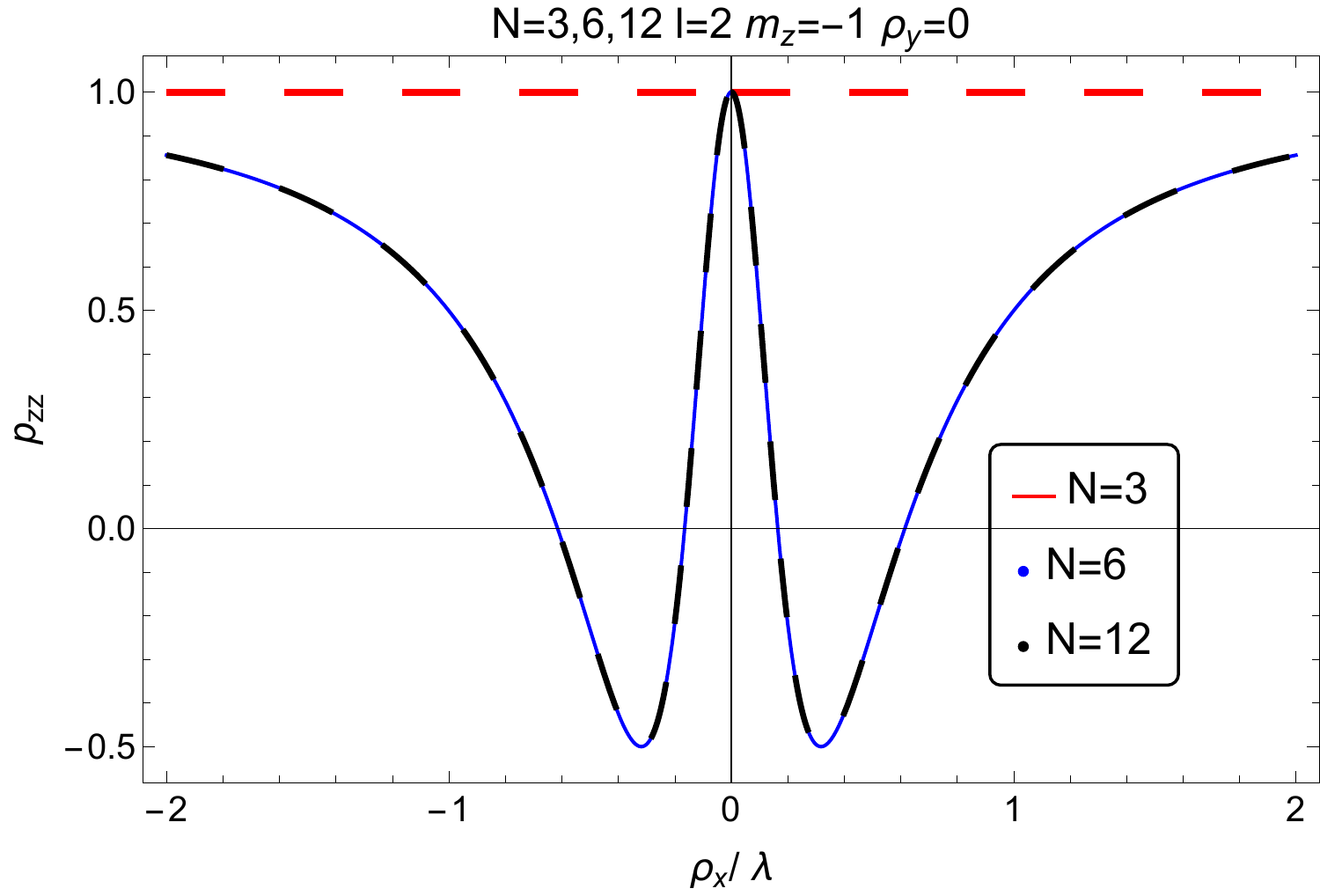}
\caption{Polarization parameter $p_{z}$ (left plot) and $p_{zz}$ (right plot) vs radial position $\rho$ for different numbers of emitting atoms $N=3,6,12$ ($l=2$ and $m_z=-1$ and $z=z_R$). Note different behavior for $N=3$ (see the text for details).}
\label{fig:d1}
\end{figure*}

The effect on polarization from varying the number of emitters $N$ can be seen in Fig.\ref{fig:d1}. It should be noted that for $l=2$ and $N=3$ the leading-order vortex contribution has a topological charge $n=-1$ (equal in sign to $m_z$), according to Eqs.(\ref{eq:d17}-\ref{eq:d18}), which results in a different behavior of polarizations in Fig.\ref{fig:d1} as opposed to large values of $N=6,12$. 

\begin{figure*}[htbp]
\centering
\includegraphics[width=0.45\textwidth]{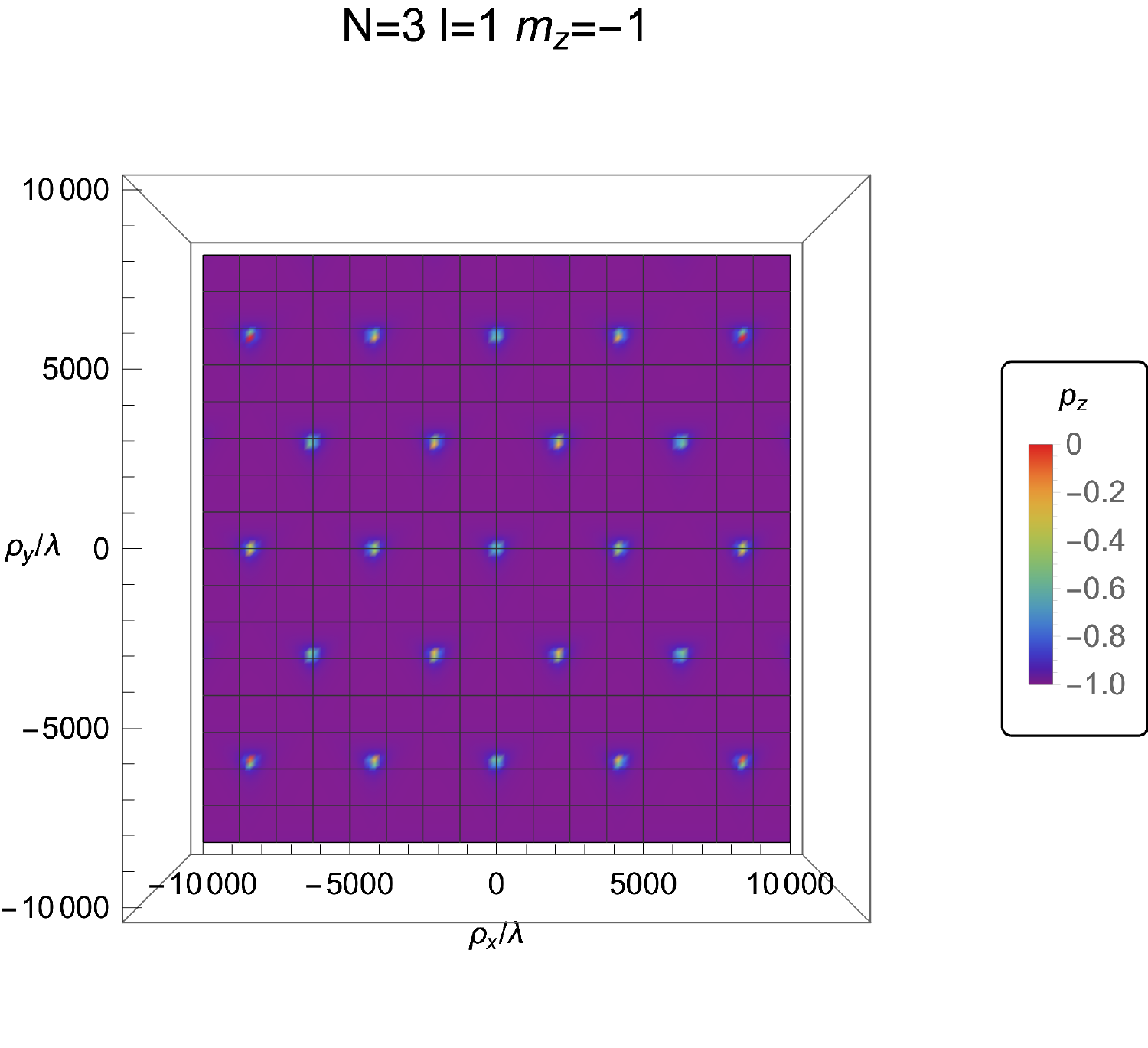}
\includegraphics[width=0.45\textwidth]{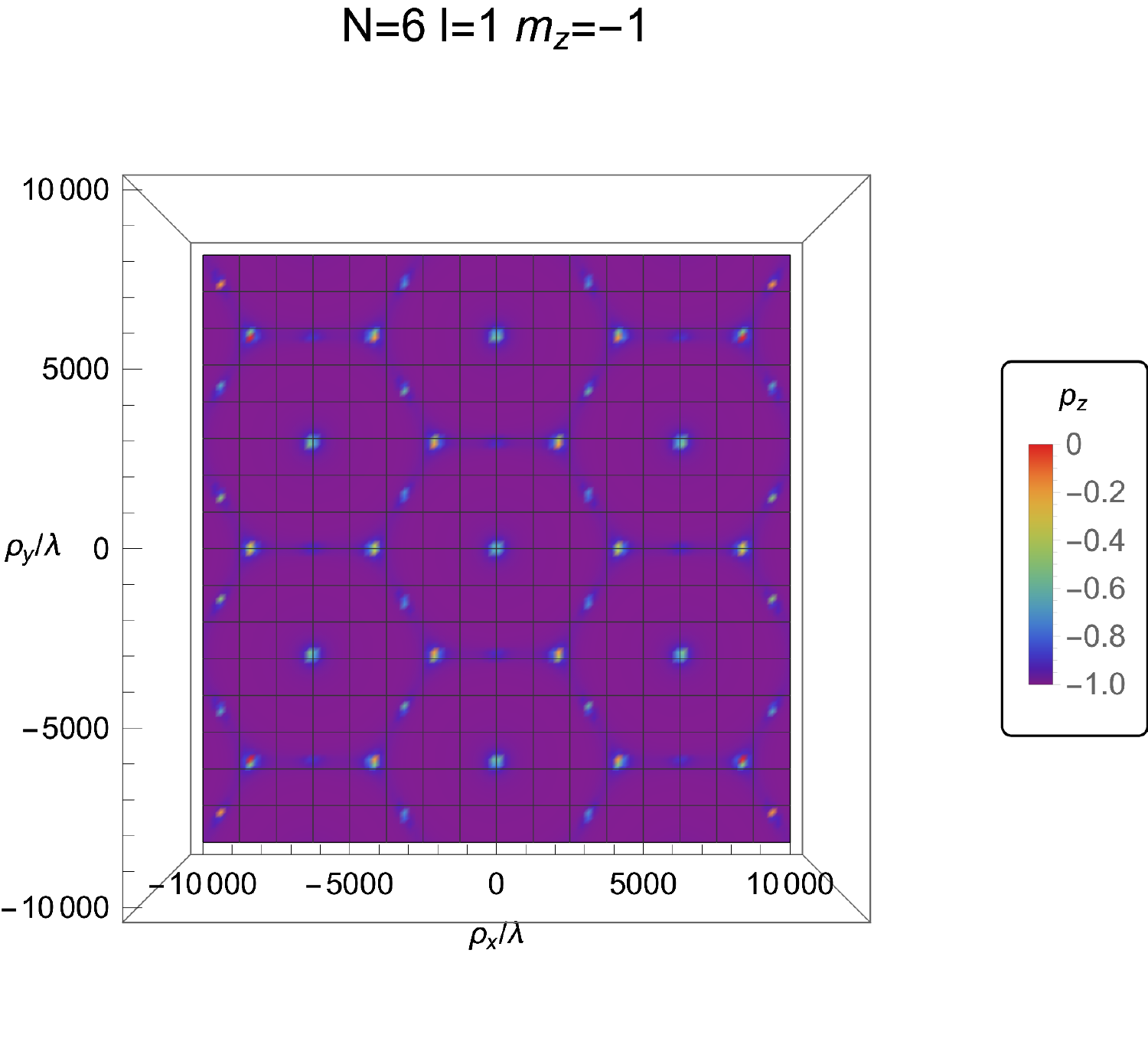}
\includegraphics[width=0.45\textwidth]{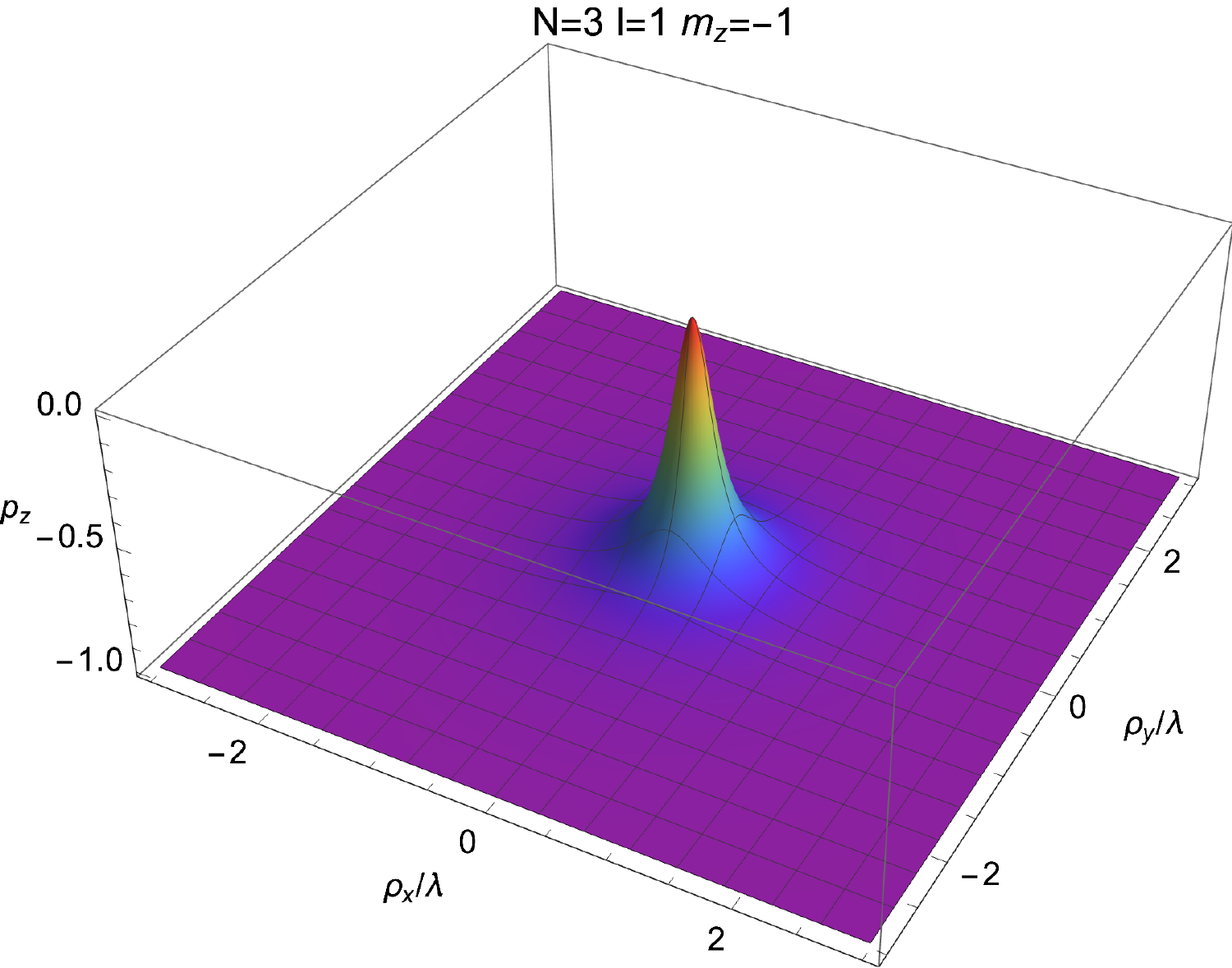}
\includegraphics[width=0.45\textwidth]{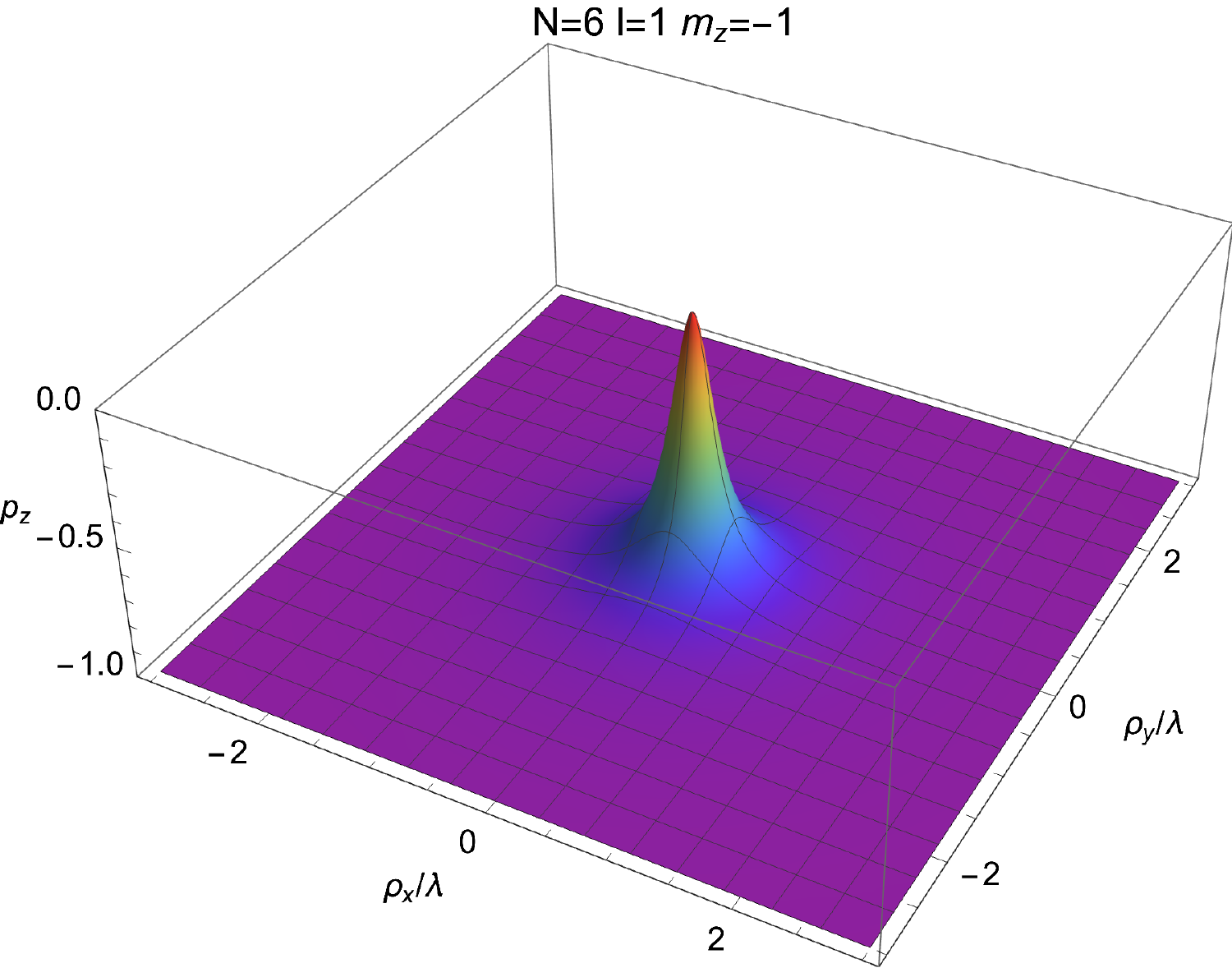}
\caption{Lattices of polarization singularities are formed for a minimal number of emitters, $N=3$ (upper left plot) and for $N=6$ (upper right plot). Close-up near $\rho=0$ is shown for $N=3$ (lower left) and $N=6$ (lower right). Here, $l=1$, $m_z=-1$ and $z=2z_R$.}
\label{fig:c4}
\end{figure*}

For sparse arrays, $N=3, 6$, polarization singularities form a lattice, as shown in Fig.\ref{fig:c4} for $p_z$. The distributions of the peaks of $p_z$ and $p_{zz}$ are closely related to the lattice-like distribution of the intensities, while spatial separation between the singularities increases with the propagation distance $z$. However, the spatial extent of an individual singularity in the vortex center remains propagation-independent and $N$-independent (once a requirement for the least number $N$ is met), as seen in bottom plots in Fig.\ref{fig:c4}. The polarization $p_{zz}$ shows similar behavior.

\section{Summary}

Let us summarize the obtained results. We considered a circular array of $N$ atoms excited into a $P$-state with a definite magnetic quantum number $m_z$ and analyzed topological properties of their coherent radiation, provided that it is controlled with a phase parameter $l$. We found that for $N\to\infty$ the resulting radiation near the array axis is described by a superposition of circularly polarized Bessel beams with different topological charges. Next, we derived a result for the total vector potential of a phased atomic array with $N$ atoms and a phase parameter $l$ expressed in terms of an infinite series of Bessel vortices.
We obtained necessary conditions for the least number $N$ required to generate vortex states with a given topological charge $l$ for all components of the vector field: $N\geq2(l+m_z)+3$. 

In analyzing the resulting vortex field properties, we applied a formalism of 3D-polarization \cite{Afanasev_2020}, with a spin-density matrix of the twisted photons defined in terms of eight polarization parameters. We demonstrated that if the spin and orbital angular momentum of the radiation are anti-aligned, then spatial variation of radiation's polarization in the plane parallel to the atomic array near the vortex axis is independent of the distance to the source. 
Presence of a longitudinal component of the field plays a significant role in this behavior. In order to fulfill Maxwell's equations, optical vortex beams have to include non-zero longitudinal components \cite{Afanasev2022non}, with the field ratios showing propagation-independent spatial features. 
Here, we found similar propagation-invariant polarization features in radiation from atomic arrays.

The developed formalism can be extended and applied in several areas, from conventional polarized phased antenna arrays to complex geometries of multi-atom Dicke states, to the transfer of quantum information between atomic ensembles.

\section*{Acknowledgements}
We thank US Army Research Office for support under Grant W911NF-19-1-0022. AA thanks Peter Zoller for useful discussions.

\bibliography{arraybib.bib}

\begin{thebibliography}{17}%
\makeatletter
\providecommand \@ifxundefined [1]{%
 \@ifx{#1\undefined}
}%
\providecommand \@ifnum [1]{%
 \ifnum #1\expandafter \@firstoftwo
 \else \expandafter \@secondoftwo
 \fi
}%
\providecommand \@ifx [1]{%
 \ifx #1\expandafter \@firstoftwo
 \else \expandafter \@secondoftwo
 \fi
}%
\providecommand \natexlab [1]{#1}%
\providecommand \enquote  [1]{``#1''}%
\providecommand \bibnamefont  [1]{#1}%
\providecommand \bibfnamefont [1]{#1}%
\providecommand \citenamefont [1]{#1}%
\providecommand \href@noop [0]{\@secondoftwo}%
\providecommand \href [0]{\begingroup \@sanitize@url \@href}%
\providecommand \@href[1]{\@@startlink{#1}\@@href}%
\providecommand \@@href[1]{\endgroup#1\@@endlink}%
\providecommand \@sanitize@url [0]{\catcode `\\12\catcode `\$12\catcode
  `\&12\catcode `\#12\catcode `\^12\catcode `\_12\catcode `\%12\relax}%
\providecommand \@@startlink[1]{}%
\providecommand \@@endlink[0]{}%
\providecommand \url  [0]{\begingroup\@sanitize@url \@url }%
\providecommand \@url [1]{\endgroup\@href {#1}{\urlprefix }}%
\providecommand \urlprefix  [0]{URL }%
\providecommand \Eprint [0]{\href }%
\providecommand \doibase [0]{http://dx.doi.org/}%
\providecommand \selectlanguage [0]{\@gobble}%
\providecommand \bibinfo  [0]{\@secondoftwo}%
\providecommand \bibfield  [0]{\@secondoftwo}%
\providecommand \translation [1]{[#1]}%
\providecommand \BibitemOpen [0]{}%
\providecommand \bibitemStop [0]{}%
\providecommand \bibitemNoStop [0]{.\EOS\space}%
\providecommand \EOS [0]{\spacefactor3000\relax}%
\providecommand \BibitemShut  [1]{\csname bibitem#1\endcsname}%
\let\auto@bib@innerbib\@empty
\bibitem [{\citenamefont {Franke-Arnold}\ and\ \citenamefont
  {Radwell}(2017)}]{franke2017light}%
  \BibitemOpen
  \bibfield  {author} {\bibinfo {author} {\bibfnamefont {S.}~\bibnamefont
  {Franke-Arnold}}\ and\ \bibinfo {author} {\bibfnamefont {N.}~\bibnamefont
  {Radwell}},\ }\href
  {https://www.optica-opn.org/home/articles/volume_28/june_2017/features/light_served_with_a_twist/?testing}
  {\bibfield  {journal} {\bibinfo  {journal} {Optics and Photonics News}\
  }\textbf {\bibinfo {volume} {28}},\ \bibinfo {pages} {28} (\bibinfo {year}
  {2017})}\BibitemShut {NoStop}%
\bibitem [{\citenamefont {Willner}(2021)}]{Willner2021oam}%
  \BibitemOpen
  \bibfield  {author} {\bibinfo {author} {\bibfnamefont {A.~E.}\ \bibnamefont
  {Willner}},\ }\href@noop {} {\bibfield  {journal} {\bibinfo  {journal}
  {Optics and Photonics News}\ }\textbf {\bibinfo {volume} {32}},\ \bibinfo
  {pages} {34} (\bibinfo {year} {2021})}\BibitemShut {NoStop}%
\bibitem [{\citenamefont {Thid{\'e}}\ and\ \citenamefont
  {Tamburini}(2021)}]{thide2021oam}%
  \BibitemOpen
  \bibfield  {author} {\bibinfo {author} {\bibfnamefont {B.}~\bibnamefont
  {Thid{\'e}}}\ and\ \bibinfo {author} {\bibfnamefont {F.}~\bibnamefont
  {Tamburini}},\ }\href@noop {} {\emph {\bibinfo {title} {OAM Radio--physical
  foundations and applications of electromagnetic orbital angular momentum in
  radio science and technology}}}\ (\bibinfo  {publisher} {Wiley Online
  Library},\ \bibinfo {year} {2021})\ pp.\ \bibinfo {pages}
  {33--95}\BibitemShut {NoStop}%
\bibitem [{\citenamefont {Wang}\ \emph {et~al.}(2009)\citenamefont {Wang},
  \citenamefont {Wang},\ and\ \citenamefont {Zhu}}]{WANG20091088}%
  \BibitemOpen
  \bibfield  {author} {\bibinfo {author} {\bibfnamefont {L.-G.}\ \bibnamefont
  {Wang}}, \bibinfo {author} {\bibfnamefont {L.-Q.}\ \bibnamefont {Wang}}, \
  and\ \bibinfo {author} {\bibfnamefont {S.-Y.}\ \bibnamefont {Zhu}},\ }\href
  {\doibase https://doi.org/10.1016/j.optcom.2008.12.004} {\bibfield  {journal}
  {\bibinfo  {journal} {Optics Communications}\ }\textbf {\bibinfo {volume}
  {282}},\ \bibinfo {pages} {1088} (\bibinfo {year} {2009})}\BibitemShut
  {NoStop}%
\bibitem [{\citenamefont {Liu}\ \emph {et~al.}(2015)\citenamefont {Liu},
  \citenamefont {Phillips}, \citenamefont {Li}, \citenamefont {Williams},
  \citenamefont {Andrews},\ and\ \citenamefont {Padgett}}]{Liu_2015}%
  \BibitemOpen
  \bibfield  {author} {\bibinfo {author} {\bibfnamefont {R.}~\bibnamefont
  {Liu}}, \bibinfo {author} {\bibfnamefont {D.~B.}\ \bibnamefont {Phillips}},
  \bibinfo {author} {\bibfnamefont {F.}~\bibnamefont {Li}}, \bibinfo {author}
  {\bibfnamefont {M.~D.}\ \bibnamefont {Williams}}, \bibinfo {author}
  {\bibfnamefont {D.~L.}\ \bibnamefont {Andrews}}, \ and\ \bibinfo {author}
  {\bibfnamefont {M.~J.}\ \bibnamefont {Padgett}},\ }\href {\doibase
  10.1088/2040-8978/17/4/045608} {\bibfield  {journal} {\bibinfo  {journal}
  {Journal of Optics}\ }\textbf {\bibinfo {volume} {17}},\ \bibinfo {pages}
  {045608} (\bibinfo {year} {2015})}\BibitemShut {NoStop}%
\bibitem [{\citenamefont {Perczel}\ \emph {et~al.}(2017)\citenamefont
  {Perczel}, \citenamefont {Borregaard}, \citenamefont {Chang}, \citenamefont
  {Pichler}, \citenamefont {Yelin}, \citenamefont {Zoller},\ and\ \citenamefont
  {Lukin}}]{perczel2017topological}%
  \BibitemOpen
  \bibfield  {author} {\bibinfo {author} {\bibfnamefont {J.}~\bibnamefont
  {Perczel}}, \bibinfo {author} {\bibfnamefont {J.}~\bibnamefont {Borregaard}},
  \bibinfo {author} {\bibfnamefont {D.~E.}\ \bibnamefont {Chang}}, \bibinfo
  {author} {\bibfnamefont {H.}~\bibnamefont {Pichler}}, \bibinfo {author}
  {\bibfnamefont {S.~F.}\ \bibnamefont {Yelin}}, \bibinfo {author}
  {\bibfnamefont {P.}~\bibnamefont {Zoller}}, \ and\ \bibinfo {author}
  {\bibfnamefont {M.~D.}\ \bibnamefont {Lukin}},\ }\href@noop {} {\bibfield
  {journal} {\bibinfo  {journal} {Phys. Rev. Lett.}\ }\textbf {\bibinfo
  {volume} {119}},\ \bibinfo {pages} {023603} (\bibinfo {year}
  {2017})}\BibitemShut {NoStop}%
\bibitem [{\citenamefont {Wang}\ \emph {et~al.}(2022)\citenamefont {Wang},
  \citenamefont {Szekerczes},\ and\ \citenamefont {Afanasev}}]{Wang_2022}%
  \BibitemOpen
  \bibfield  {author} {\bibinfo {author} {\bibfnamefont {H.}~\bibnamefont
  {Wang}}, \bibinfo {author} {\bibfnamefont {K.}~\bibnamefont {Szekerczes}}, \
  and\ \bibinfo {author} {\bibfnamefont {A.}~\bibnamefont {Afanasev}},\ }\href
  {\doibase 10.1088/2399-6528/ac5089} {\bibfield  {journal} {\bibinfo
  {journal} {Journal of Physics Communications}\ }\textbf {\bibinfo {volume}
  {6}},\ \bibinfo {pages} {025005} (\bibinfo {year} {2022})}\BibitemShut
  {NoStop}%
\bibitem [{\citenamefont {Jentschura}\ and\ \citenamefont
  {Serbo}(2011)}]{Jentschura11}%
  \BibitemOpen
  \bibfield  {author} {\bibinfo {author} {\bibfnamefont {U.~D.}\ \bibnamefont
  {Jentschura}}\ and\ \bibinfo {author} {\bibfnamefont {V.~G.}\ \bibnamefont
  {Serbo}},\ }\href {\doibase 10.1103/PhysRevLett.106.013001} {\bibfield
  {journal} {\bibinfo  {journal} {Phys. Rev. Lett.}\ }\textbf {\bibinfo
  {volume} {106}},\ \bibinfo {pages} {013001} (\bibinfo {year}
  {2011})}\BibitemShut {NoStop}%
\bibitem [{\citenamefont {Afanasev}\ \emph {et~al.}(2013)\citenamefont
  {Afanasev}, \citenamefont {Carlson},\ and\ \citenamefont
  {Mukherjee}}]{PhysRevA.88.033841}%
  \BibitemOpen
  \bibfield  {author} {\bibinfo {author} {\bibfnamefont {A.}~\bibnamefont
  {Afanasev}}, \bibinfo {author} {\bibfnamefont {C.~E.}\ \bibnamefont
  {Carlson}}, \ and\ \bibinfo {author} {\bibfnamefont {A.}~\bibnamefont
  {Mukherjee}},\ }\href {\doibase 10.1103/PhysRevA.88.033841} {\bibfield
  {journal} {\bibinfo  {journal} {Phys. Rev. A}\ }\textbf {\bibinfo {volume}
  {88}},\ \bibinfo {pages} {033841} (\bibinfo {year} {2013})}\BibitemShut
  {NoStop}%
\bibitem [{\citenamefont {Sobelman}(2012)}]{sobelman2012atomic}%
  \BibitemOpen
  \bibfield  {author} {\bibinfo {author} {\bibfnamefont {I.~I.}\ \bibnamefont
  {Sobelman}},\ }\href@noop {} {\emph {\bibinfo {title} {Atomic spectra and
  radiative transitions}}},\ Vol.~\bibinfo {volume} {12}\ (\bibinfo
  {publisher} {Springer Science \& Business Media},\ \bibinfo {year}
  {2012})\BibitemShut {NoStop}%
\bibitem [{\citenamefont {Sheppard}(2014)}]{PhysRevA.90.023809}%
  \BibitemOpen
  \bibfield  {author} {\bibinfo {author} {\bibfnamefont {C.~J.~R.}\
  \bibnamefont {Sheppard}},\ }\href {\doibase 10.1103/PhysRevA.90.023809}
  {\bibfield  {journal} {\bibinfo  {journal} {Phys. Rev. A}\ }\textbf {\bibinfo
  {volume} {90}},\ \bibinfo {pages} {023809} (\bibinfo {year}
  {2014})}\BibitemShut {NoStop}%
\bibitem [{\citenamefont {Eismann}\ \emph {et~al.}(2021)\citenamefont
  {Eismann}, \citenamefont {Nicholls}, \citenamefont {Roth}, \citenamefont
  {Alonso}, \citenamefont {Banzer}, \citenamefont {Rodr{\'\i}guez-Fortu{\~n}o},
  \citenamefont {Zayats}, \citenamefont {Nori},\ and\ \citenamefont
  {Bliokh}}]{Eismann:2021te}%
  \BibitemOpen
  \bibfield  {author} {\bibinfo {author} {\bibfnamefont {J.~S.}\ \bibnamefont
  {Eismann}}, \bibinfo {author} {\bibfnamefont {L.~H.}\ \bibnamefont
  {Nicholls}}, \bibinfo {author} {\bibfnamefont {D.~J.}\ \bibnamefont {Roth}},
  \bibinfo {author} {\bibfnamefont {M.~A.}\ \bibnamefont {Alonso}}, \bibinfo
  {author} {\bibfnamefont {P.}~\bibnamefont {Banzer}}, \bibinfo {author}
  {\bibfnamefont {F.~J.}\ \bibnamefont {Rodr{\'\i}guez-Fortu{\~n}o}}, \bibinfo
  {author} {\bibfnamefont {A.~V.}\ \bibnamefont {Zayats}}, \bibinfo {author}
  {\bibfnamefont {F.}~\bibnamefont {Nori}}, \ and\ \bibinfo {author}
  {\bibfnamefont {K.~Y.}\ \bibnamefont {Bliokh}},\ }\href {\doibase
  http://dx.doi.org/10.1038/s41566-020-00733-3} {\bibfield  {journal} {\bibinfo
   {journal} {Nature Photonics}\ }\textbf {\bibinfo {volume} {15}},\ \bibinfo
  {pages} {156} (\bibinfo {year} {2021})}\BibitemShut {NoStop}%
\bibitem [{\citenamefont {Afanasev}\ \emph {et~al.}(2020)\citenamefont
  {Afanasev}, \citenamefont {Carlson},\ and\ \citenamefont
  {Wang}}]{Afanasev_2020}%
  \BibitemOpen
  \bibfield  {author} {\bibinfo {author} {\bibfnamefont {A.}~\bibnamefont
  {Afanasev}}, \bibinfo {author} {\bibfnamefont {C.~E.}\ \bibnamefont
  {Carlson}}, \ and\ \bibinfo {author} {\bibfnamefont {H.}~\bibnamefont
  {Wang}},\ }\href {\doibase 10.1088/2040-8986/ab8288} {\bibfield  {journal}
  {\bibinfo  {journal} {Journal of Optics}\ }\textbf {\bibinfo {volume} {22}},\
  \bibinfo {pages} {054001} (\bibinfo {year} {2020})}\BibitemShut {NoStop}%
\bibitem [{\citenamefont {Ohlsen}(1972)}]{Ohlsen_1972}%
  \BibitemOpen
  \bibfield  {author} {\bibinfo {author} {\bibfnamefont {G.~G.}\ \bibnamefont
  {Ohlsen}},\ }\href {\doibase 10.1088/0034-4885/35/2/305} {\bibfield
  {journal} {\bibinfo  {journal} {Reports on Progress in Physics}\ }\textbf
  {\bibinfo {volume} {35}},\ \bibinfo {pages} {717} (\bibinfo {year}
  {1972})}\BibitemShut {NoStop}%
\bibitem [{\citenamefont {Varshalovich}\ \emph {et~al.}(1988)\citenamefont
  {Varshalovich}, \citenamefont {Moskalev},\ and\ \citenamefont
  {Khersonskii}}]{varshalovich1988quantum}%
  \BibitemOpen
  \bibfield  {author} {\bibinfo {author} {\bibfnamefont {D.~A.}\ \bibnamefont
  {Varshalovich}}, \bibinfo {author} {\bibfnamefont {A.~N.}\ \bibnamefont
  {Moskalev}}, \ and\ \bibinfo {author} {\bibfnamefont {V.~K.}\ \bibnamefont
  {Khersonskii}},\ }\href@noop {} {\emph {\bibinfo {title} {Quantum theory of
  angular momentum}}}\ (\bibinfo  {publisher} {World Scientific},\ \bibinfo
  {year} {1988})\BibitemShut {NoStop}%
\bibitem [{\citenamefont {Carozzi}\ \emph {et~al.}(2000)\citenamefont
  {Carozzi}, \citenamefont {Karlsson},\ and\ \citenamefont
  {Bergman}}]{carozzi2000}%
  \BibitemOpen
  \bibfield  {author} {\bibinfo {author} {\bibfnamefont {T.}~\bibnamefont
  {Carozzi}}, \bibinfo {author} {\bibfnamefont {R.}~\bibnamefont {Karlsson}}, \
  and\ \bibinfo {author} {\bibfnamefont {J.}~\bibnamefont {Bergman}},\ }\href
  {\doibase 10.1103/PhysRevE.61.2024} {\bibfield  {journal} {\bibinfo
  {journal} {Phys. Rev. E}\ }\textbf {\bibinfo {volume} {61}},\ \bibinfo
  {pages} {2024} (\bibinfo {year} {2000})}\BibitemShut {NoStop}%
\bibitem [{\citenamefont {Afanasev}\ \emph {et~al.}(2022)\citenamefont
  {Afanasev}, \citenamefont {Kingsley-Smith}, \citenamefont
  {Rodr{\'\i}guez-Fortu{\~n}o},\ and\ \citenamefont
  {Zayats}}]{Afanasev2022non}%
  \BibitemOpen
  \bibfield  {author} {\bibinfo {author} {\bibfnamefont {A.}~\bibnamefont
  {Afanasev}}, \bibinfo {author} {\bibfnamefont {J.~J.}\ \bibnamefont
  {Kingsley-Smith}}, \bibinfo {author} {\bibfnamefont {F.~J.}\ \bibnamefont
  {Rodr{\'\i}guez-Fortu{\~n}o}}, \ and\ \bibinfo {author} {\bibfnamefont
  {A.~V.}\ \bibnamefont {Zayats}},\ }\href {https://arxiv.org/abs/2208.08833}
  {\bibfield  {journal} {\bibinfo  {journal} {arXiv preprint arXiv:2208.08833}\
  } (\bibinfo {year} {2022})}\BibitemShut {NoStop}%
\end{thebibliography}%



\end{document}